\documentclass[10pt,journal,compsoc]{IEEEtran}
\usepackage{amsmath,amsfonts}
\usepackage{algorithmic}
\usepackage{algorithm}
\usepackage{array}
\usepackage[caption=false,font=normalsize,labelfont=sf,textfont=sf]{subfig}
\usepackage{textcomp}
\usepackage{stfloats}
\usepackage{url}
\usepackage{verbatim}
\usepackage{graphicx}
\usepackage{cite}
\usepackage{amsmath,amssymb,amsfonts}
\usepackage{algorithm}
\usepackage{algorithmic}
\usepackage{graphicx}
\usepackage{subfloat}
\usepackage{textcomp}
\def\BibTeX{{\rm B\kern-.05em{\sc i\kern-.025em b}\kern-.08em
    T\kern-.1667em\lower.7ex\hbox{E}\kern-.125emX}}

\usepackage{xcolor}
\usepackage{multirow}
\usepackage{booktabs}
\usepackage{makecell}
\usepackage{tabu}
\usepackage{booktabs}
\usepackage{longtable}
\usepackage{fontawesome}
\usepackage{url}
\usepackage{eqparbox}

\begin{document}
\title{SGBA: A Stealthy Scapegoat Backdoor Attack against Deep Neural Networks}

This work has been submitted to the IEEE for possible publication. Copyright may be transferred without notice, after which this version may no longer be accessible.

\author{Ying~He, Zhili~Shen, Chang~Xia, Jingyu~Hua, Wei~Tong, and Sheng~Zhong
        % <-this % stops a space
\IEEEcompsocitemizethanks{\IEEEcompsocthanksitem Y. He, Z. Shen and C. Xia are with the Computer Science and Technology Department, Nanjing University, Nanjing 210023, China. E-mail: guapi7878@gmail.com; zhilishen@163.com; changxia656569@gmail.com.
\IEEEcompsocthanksitem J. Hua, W. Tong and S. Zhong are with the State Key Laboratory for Novel Software Technology, Nanjing University, Nanjing 210023, China, and also with
the Computer Science and Technology Department, Nanjing University,
Nanjing 210023, China. E-mail: huajingyu@nju.edu.cn; weitong@outlook.com; zhongsheng@nju.edu.cn.}% <-this % stops a space
% \thanks{Manuscript received April 19, 2005; revised August 26, 2015.}
\thanks{This work was supported in part by NSFC-61972195, the Leading-edge Technology Program of Jiangsu NSF (No.BK20202001), NSFC-61872179, NSFC-61872176 and NSFC-62002159.}% <-this % stops a space
\thanks{(Corresponding authors: Jingyu Hua and Wei Tong.)}% <-this % stops a space
}

\markboth{IEEE TRANSACTIONS ON DEPENDABLE AND SECURE COMPUTING,~Vol.~14, No.~8, August~2021}%
{Shell \MakeLowercase{\textit{et al.}}: SGBA: A Stealthy Scapegoat Backdoor Attack against Deep Neural Networks}

\IEEEpubid{0000--0000/00\$00.00~\copyright~2021 IEEE}

\IEEEtitleabstractindextext{%
\begin{abstract}
Outsourced deep neural networks have been demonstrated to suffer from patch-based trojan attacks, in which an adversary poisons the training sets to inject a backdoor in the obtained model so that regular inputs can be  still labeled correctly while those carrying a specific trigger are falsely given a target label. Due to the severity of such attacks, many backdoor detection and containment systems have recently, been proposed for deep neural networks. One major category among them are various model inspection schemes, which hope to detect backdoors before deploying models from non-trusted third-parties. In this paper, we show that such state-of-the-art schemes can be defeated by a so-called Scapegoat Backdoor Attack, which introduces a benign scapegoat trigger in data poisoning to prevent the defender from reversing the real abnormal trigger. In addition, it confines the values of network parameters within the same variances of those from clean model during training, which further significantly enhances the difficulty of the defender to learn the differences between legal and illegal models through machine-learning approaches. Our experiments on 3 popular datasets show that it can escape detection by all five state-of-the-art model inspection schemes. Moreover, this attack brings almost no side-effects on the attack effectiveness and guarantees the universal feature of the trigger compared with original patch-based trojan attacks.
\end{abstract}

\begin{IEEEkeywords}
backdoor attack, deep neural network, scapegoat, data poisoning, weight limitation
\end{IEEEkeywords}}

\maketitle

\IEEEdisplaynontitleabstractindextext

\IEEEpeerreviewmaketitle

\section{Introduction}
\IEEEPARstart{D}{eep} \emph{Neural Networks} (DNN) show a strong learning ability due to their deep network architectures. As a result, they haven been widely applied in many tasks such as image classification and recognition \cite{b41}, speech recognition and machine translation \cite{b43} as well as text classification \cite{b42}. DNN usually needs a large number of samples to train the model. Be it for collecting or labeling data, large resources are consumed. Besides, training a model also requires high computing abilities of both CPUs and GPUs. Therefore, many users choose to reuse a pre-trained model from third-parties. 

As not all the third-parties are trustworthy, such pre-trained models are demonstrated to suffer from \emph{Patch-based Trojan Attacks} (PTAs) \cite{b1,b2,b3}, in which a malicious model trainer may inject a backdoor in the model so that regular inputs are still labeled correctly while those carrying a specific trigger will always be given a false target label regardless of their true ones. Here, a trigger is a small but arbitrary pattern (e.g., a small square) specified by the attacker. The backdoors could be planted by poisoning the training data with a partially mislabeled data stamped with the trigger. 

\IEEEpubidadjcol
As trojan models may raise serious security threats, they have received extensive attentions and many containment proposals have been discussed in the literature \cite{b6,b7,b9,b10,b11,b12,b13,b14,b15,b16,b17,b18,b35,b38}. The most promising strategy is to deploy a model inspection scheme that aims to detect the anomalies due to the existence of backdoors before really applying the model in practice. According to existing works, the SOTA DNN model inspection schemes against backdoors can be briefly divided into two categories \cite{b36}. The first category \cite{b14,b9,b13,b18} attempts to reverse engineer a trigger pattern for each class that can cause the misclassifications of all the inputs carrying it to the corresponding class, through various optimization-based methods. As the backdoor actually creates a `shortcut' to the target class, they assume the reversed trigger of the target would be abnormal compared with those of normal classes and thus be detected. However, simple search strategies in the reverse engineering can be stuck in sub-optimums. Although some of them enhance the search object restriction, we will prove it still can be misled to sub-optimums in the paper. The second category \cite{b6,b35,b38} employs the machine learning technique to recognize trojaned models. In particular, they first build and label both a large number of clean and trojaned models, and then take them as a training set to learn a binary classifier that can automatically differentiate benign and trojaned models with a high confidence. These defense schemes rely on the differences learned from benign models and backdoored models of the training set. We will show that once the malicious model is close enough to benign models, the learning ability of the detection classifier would reduce.

In this paper, we propose a stealthy \emph{Scapegoat Backdoor Attack} (SGBA), which can evade detection by all existing model inspection schemes while maintaining the original concept of PTAs. Note that, recently, some researchers have proposed some novel DNN trojan attacks that can really fool these schemes \cite{b5,b40,b25,b24}. Unfortunately, most of them more or less change the original definition of PTAs. For instance, in the composition attack \cite{b5}, the trigger can no longer be an arbitrary pattern, but should be benign subjects. Moreover, in both this attack and the TaCT attack \cite{b46}, the universal of the effective domain of a single trigger decreases. In the original PTAs, a trigger is universal and can affect all the inputs carrying it. We think the loss of such a universal feature brings a non-negligible sacrifice to the attack feasibility.

In particular, to defeat the trigger reverse-engineering schemes, our attack introduces a benign scapegoat trigger in data poisoning to hide the real trigger, which is constituted by two separated but interdependent patterns (both are small and arbitrary shapes) specified by the adversary. It places a loose restriction on the relative positions of the scapegoat and the real trigger on the input pixel space, demonstrating that it can misguide the existing trigger reverse-engineering methods to just reverse the scapegoat. In our proposal, a scapegoat is actually a benign trigger reversed from a clean model and thus will not raise any alarms. Moreover, during the training of the malicious model, we further confine the values of network parameters within the same variances as those of a clean model. According to our analysis, the weight variances of malicious models is much bigger than those of benign models especially in the latter layers, which might be because the malicious model has to make some weights unusually higher in order to react to the trigger. Our restriction can thus significantly reduces the anomaly of the obtained model compared with the clean ones, which obviously largely enhances the learning difficulty of ML-based detection mechanisms to differentiate them. In addition, our experiments prove that the proposed attack can still defeat such ML-based inspection schemes even when they know our attack and include our malicious models in the training set.

In summary, we make the following contributions:
\begin{itemize}
    \item We propose a novel stealthy patch-based trojan attack against DNN in the training outsourcing scenario. It does not change the original concept of PTAs and can guarantee the universal effective domain of the trigger. Moreover, it introduces a scapegoat trigger to hide the real one and restricts the deviation of the model parameters from a clean model during training.
    \item We analyze and demonstrate the effectiveness of the proposed attack against all the existing model inspection mechanisms including those based on trigger reverse-engineering and those based on machine learning. 
    \item We apply our attack to three popular datasets of image classification including MNIST \cite{b20}, CIFAR10 \cite{b22} and GTSRB \cite{b21}. The experimental results demonstrate that the side-effect of SGBA on classification accuracy is negligible. The average detection rates of SGBA significantly decrease from more than 90\% to less than 10\% compared with  BadNets \cite{b1} (e.g. the most basic PTA) while SGBA keeps the original aggressivity at the same time. Besides, we prove that our attack can be extrapolated to bigger datasets like ImageNet \cite{b39}.
\end{itemize}

\section{background}\label{bg}
  It is necessary to introduce basic backdoor attacks and mainstream model inspection schemes before explaining  our attack scheme.

  \subsection{Basic Backdoor Attacks}
\begin{figure}[htbp]
      \centering
      \includegraphics[width=3.4in]{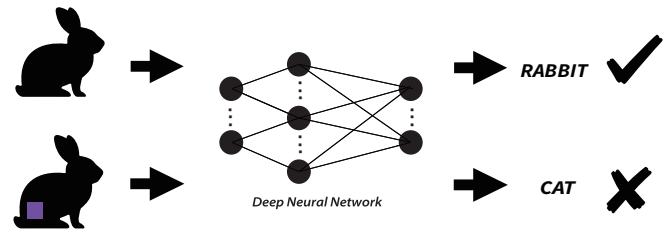}
      \caption{A simple example of backdoor attacks.}
      \label{fig1}
  \end{figure}

  As artificial flaws hidden in the system, backdoors can leak information to the attacker or help the latter obtain access that does not belong to him or her. Backdoor attacks in DNN means there is a backdoor placed in the target model by an attacker. It is invisible when the model works on a normal dataset while can be triggered by the inputs with a special pattern, which is designed by the attacker, attached. Fig.\ref{fig1} demonstrates a simple example of backdoor attacks. The model identifies the coming input as `rabbit' in a normal situation which is correct and conforms to the design rule. When the same input comes with a square sticker, the model identifies it as `cat' which is absolutely wrong. It is different from common-cause recognition failures of model learning capacity which tend to appear randomly. It's an artificial misleading recognition where inputs with the sticker tend to be identified as the same result that is usually designed by the attacker.
  
\begin{figure*}[htbp]
      \centering
      \includegraphics[width=7.0in]{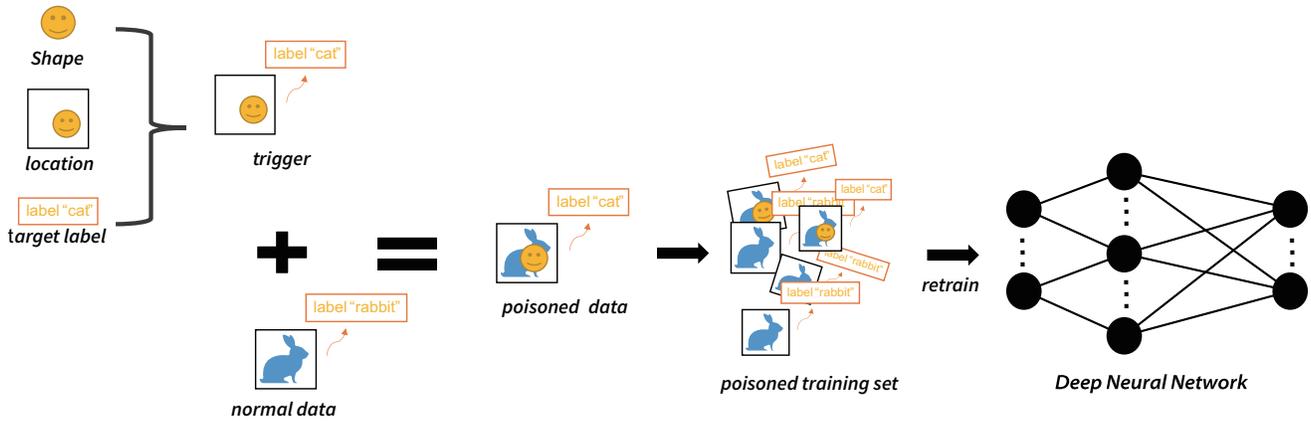}
      \caption{Poisoning and training workflow of PTAs.}
      \label{fig2}
  \end{figure*}

  The basic manipulation of backdoor attacks is to adjust the weights of the target model by data poisoning with a special trigger injected into the training set. The trigger is designed by the attacker and often has its fixed properties like shape and location in the field of image classification. This basic type of backdoor attacks is what we call PTAs. The workflow of PTAs is demonstrated in Fig.\ref{fig2}. BadNets proposed by Gu et al. \cite{b1} is the most basic PTA. What needs to be noted is that the backdoor attack can be applied into many other fields like speech recognition besides image classification. But image classification is still the most widely discussed topic in the field of backdoor attacks and also what we are only concerned with in this paper. 
  
  Some popular and ingenious extension of backdoor attacks using data poisoning have been proposed in the past few years. Gu et al. \cite{b1} propose a single-target backdoor attack and an all-to-all attack in 2017 where the single-target attack means that there is only one backdoor in the target model that misleads the result into one fixed label and the all-to-all attack means that every class of the target model is correspondingly misled to another class one-to-one. Chen et al. \cite{b3} mention a blended injection attack which uses a background image as the trigger to poison the training set by mixing the clean data with the trigger according to a certain proportion assigned by the attacker instead of replacing part of the clean data with the trigger. Another way to deploy backdoor attacks is to adjust the weights of the model directly. Liu et al. \cite{b2} propose an attack that selects a set of neurons that can be extremely activated by the trigger and strengthens the connection between the neurons and the trigger. These attacks work well when they were first proposed and have now been limited since defense schemes like trigger reverse-engineering schemes and ML-based schemes have been put forward. From the perspective of attackers, more complicated and strong attacks need to be designed.

  \subsection{Mainstream Model Inspection Schemes}
  Under the concerned scenarios, the existing model inspection schemes can be divided into two types: The trigger reverse-engineering schemes and the ML-based schemes \cite{b36}. The first type of schemes usually generates reversed triggers for every class and compares their features to find out if there is an anomaly. The second type of schemes makes use of machine learning technologies to create a black-box classifier to learn the differences between benign and trojan models. The features extracted by the trigger reverse-engineering schemes are always concrete (e.g., the size of the reversed pattern) while the differences learned by the ML-based schemes are unexplainable. The SOTA detection methods for each type are demonstrated as follows.

\begin{figure}[htbp]
      \centering
      \includegraphics[width=3.4in]{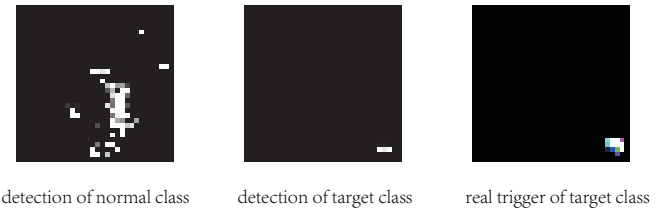}
      \caption{A simple example of Neural Cleanse's reverse results.}
      \label{fig4}
  \end{figure}

  Neural Cleanse \cite{b14} is the first and most representative of the trigger reverse-engineering schemes that detects the abnormal class from normal classes. It computes the optimal reversed triggers to convert all inputs to each target label and then checks if any trigger's $L1$ norm is significantly smaller than the others as the backdoor indicator. This is due to the fact that they thought a `shortcut' was created by the backdoor from normal classes to the target class. The `shortcut' can be reflected by the sizes of the reversed triggers (i.e., $L1$ norms).
%   It finds that the path that transforms all samples, despite any classes, into the class which has a backdoor in it is shorter than the path that transforms all samples into any of the other normal classes without a backdoor. They use the perturbation to describe such a path. For example, there are $k$ classes, a target model $M$, an input $m_i$ belonging to the $i$th class. $M(m_i)$ means the result class that model $M$ thinks the input $m_i$ belong to. And $\delta$ represents the perturbation added to the images.
%   \begin{equation}
%       M(m_i + \delta) = j
%       \label{eq1}
%   \end{equation}
%   When Eq. \eqref{eq1} is true, it means that the perturbation $\delta$ added to the input $m_i$ belonging to the $i$th class can make model $M$ take $m_i$ as the sample of the $j$th class. Now if there exists a $\delta$ that can make Eq. \eqref{eq1} true for every $m_i$, the perturbation is called a path from the $i$th class to the $j$th class. Furthermore, if there exists a $\delta$ that can make Eq. \eqref{eq1} true for every $i \in [0,k)$, the perturbation is called a path from regions of the global space to the $j$th class and we take it as $\delta_j$. When the target model has no backdoor, the value of $\delta_j$ is similar for $j \in [0,k)$. However, when the target model has a backdoor in a class, for example, the $t$th class, the value of $\delta_t$ is much smaller than others. And the $\delta_t$ is what we previous called a `shortcut'. 
  This kind of difference between the target class and normal classes in an infected model is what Neural Cleanse uses to detect whether backdoors exist or not. Fig.\ref{fig4} is a simple example of the reverse result of Neural Cleanse. It shows that the reversed trigger of a normal class is much bigger than that of the target class. At the same time, the reversed trigger of the target class is similar to the real trigger used by the attacker. 
  
  TABOR \cite{b9} improves the regularization terms of the reverse-engineering goal on the basis of Neural Cleanse to narrow down the margin of search error. They think the reverse results of Neural Cleanse are unstable because they can easily fall into the sub-optimal solutions. TABOR takes more strict limitation to the size, smoothness, location and redundancy of the reversed trigger to excludes some unexpected items which are too big, too scattered, overlaying or blocking key features of the original images. For each target it designs a corresponding regularization term to the objective function. Such improvements help to filter some sub-optimums during searching process.
  
  ABS \cite{b18} has a different optimization method from Neural Cleanse and TABOR, adopting neurons stimulation analysis to optimize the reverse engineering. It stimulates neurons from different extents and adds the one into candidate collections once it generates an extremely high activation with some stimuli. ABS scans random layers of neural networks and identifies all candidate anomaly neurons first. Then it reverses triggers for every candidate neuron. If there is a reversed trigger that can make the corresponding candidate neuron always highly activated on the clean training set, the target model is classified as a trojaned model. Compared with Neural Cleanse and TABOR, ABS only needs a small clean dataset and costs less time.

\begin{figure}[htbp]
      \centering
      \includegraphics[width=3.4in]{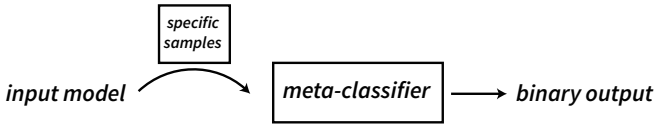}
      \caption{Simple workflow of Meta Neural Trojaned Model Detection.}
      \label{fig13}
  \end{figure}

  \emph{Meta Neural Trojaned Model Detection} (MNTD) \cite{b6} is the main outstanding of the ML-based schemes that are detecting benign and malicious models with machine learning technologies. It generates a large number of benign and trojaned models called as `shadow models' to train a binary classifier. Since the strategy of taking the whole networks of shadow models as the training set does not possess a good performance, it uses special input samples called the query set to extract feature vectors from shadow models and use them to train the classifier model. The binary result of the classifier indicates whether the target model is backdoored or not. Fig.\ref{fig13} shows the simple workflow of MNTD. Besides, to resist the adaptive attacks, MNTD proposes a robust scheme which uses random parameters for the binary classifier and only trains the query set.

  These defense schemes limit attackers from specific and abstract aspects. Attackers now needs to face more strict challenges.

\section{threat model}\label{threat}
    \subsection{Adversarial Capabilities}
    In this paper, we assume that the attacker has the full control of the training workflow of the target DNN. She/he can also access and modify the training set. We think such an assumption is reasonable when the user outsources the training to the attacker or directly downloads and fine-tunes an off-the-shelf model from a third-party market. 
    
    \subsection{Defender's Capabilities}
    We suppose that the defender has a clean dataset (below $10\%$ of the clean training set) that can be used to test the performance of the model. She/he has a full access to all the parameters and weights of the target model, but has no way to interfere with the training process performed by the model provider, i.e., the attacker in this paper.
    
    \subsection{Adversarial Goal}
    The objective of the adversary is to produce a model $F(\cdot)$ containing a stealthy patch-based backdoor by poisoning the training set. Note that although there are some other kinds of backdoors \cite{b47,b48,b49} like human invisible noise \cite{b49} that can play an important role in some aspects, we still choose patch-based backdoor because it is easier to be extended to physical attacks. Here, a so-called stealthy patch-based backdoor satisfies the following requirements:
    
    (1) It should act like a general patch-backdoor in BadNets \cite{b1} : for any input $x\in \chi$, once stamped with a trigger pattern $\tau$, they will be misclassified to the target class $C_t$ chosen by the attacker, i.e., $F(x+\tau)=C_t$. Otherwise, they will be correctly labeled. i.e., $F(x)=C_x$, where $C_x$ is the true label of $x$. In other words, we should guarantee the universal feature of the trigger. In addition, it should be a small but arbitrary pattern (e.g., a small square) specified by the attacker. 
    
    (2) It can escape from being captured by the existing model inspection mechanisms, which include two types: the first one are trigger reverse-engineering schemes (e.g., Neural Cleanse and its improvements), and the second one are ML-based schemes (e.g., MNTD).

\section{The proposed attack}\label{idea}
  We propose a new stealthy scapegoat backdoor attack which can evade the mainstream model inspection schemes. As mentioned earlier, these schemes can be briefly divided into two types: the trigger reverse-engineering schemes and the ML-based schemes \cite{b36}. In this section, we will first describe our basic methods to bypass them, respectively, and then integrate them to compose a complete robust and stealthy backdoor attack scheme.
  
  \subsection{How to bypass the trigger reverse-engineering schemes?}\label{sec_rev}
  \begin{figure}[htbp]
    \centering
    \includegraphics[width=0.95\linewidth]{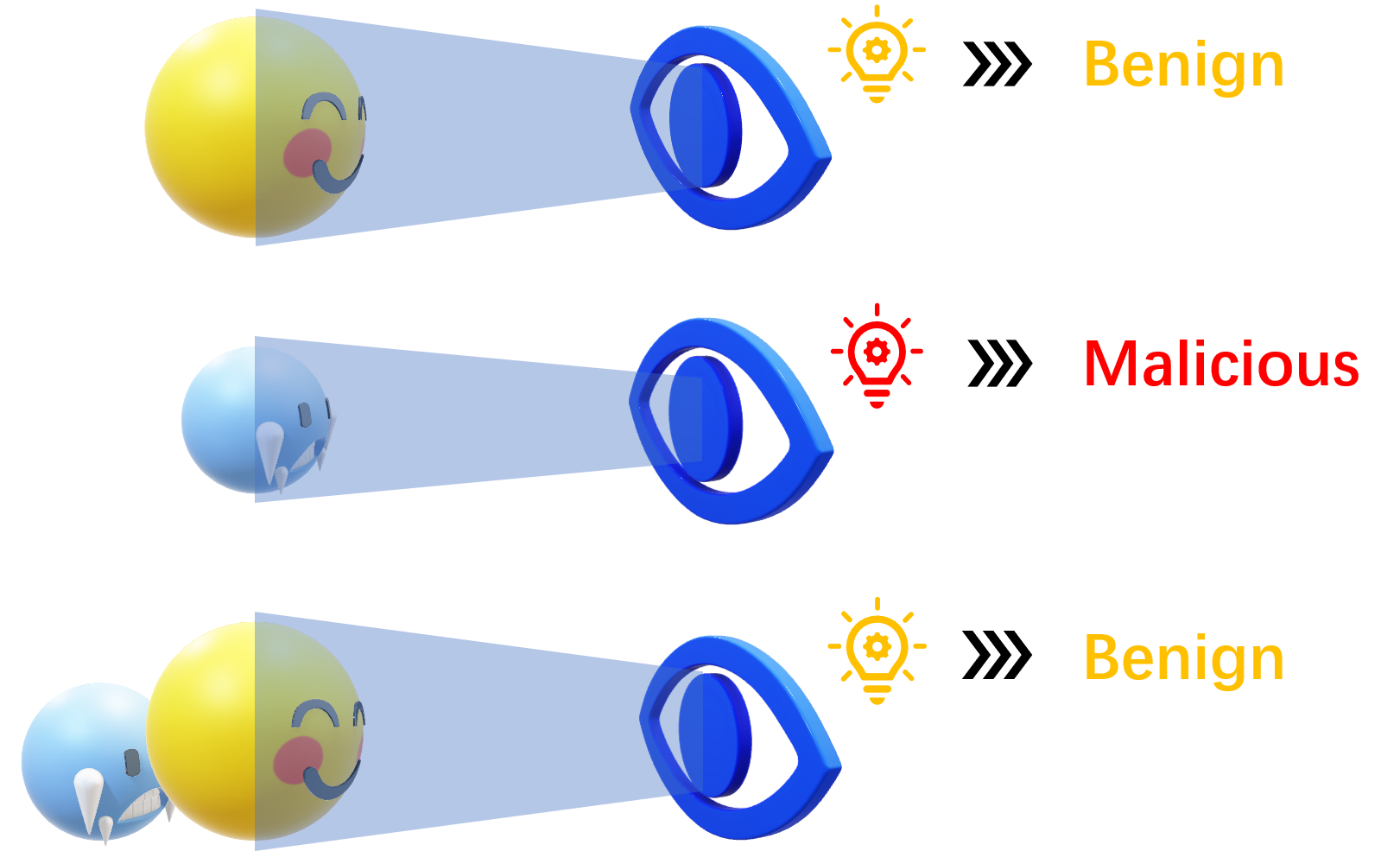}
    \caption{Basic idea of the scapegoat attack.}
    \label{new_1}
  \end{figure}
  The trigger reverse-engineering schemes are those that attempt to reverse engineer the triggers to mis-classify all the inputs to each inspected class and then determine the malicious nature of the model by comparing the reversed triggers. Note that for a clean model these schemes can still reverse a trigger for each class because in the extreme case the adversary could take an image directly from the target class as a trigger as long as it can fully cover the original inputs. We call such triggers of clean models \textit{benign triggers}, the sizes of which are considered to be close in the same model. For a malicious model with a backdoor, however, these schemes consider that there should be an abnormal reversed trigger that is much more compacted than others because the backdoor creates a `shortcut' from other classes to the target. This abnormal one is actually a close approximation of the real trigger created for the backdoor. The above is the common idea of the schemes in this category and their major difference lies in the optimization methods for trigger reverse engineering. In the following paragraphs, we first propose our attack scheme and then analyze its effectiveness against different trigger reverse-engineering schemes.
  
  \begin{figure*}[hbp]
      \centering
      \includegraphics[width=1.3\linewidth]{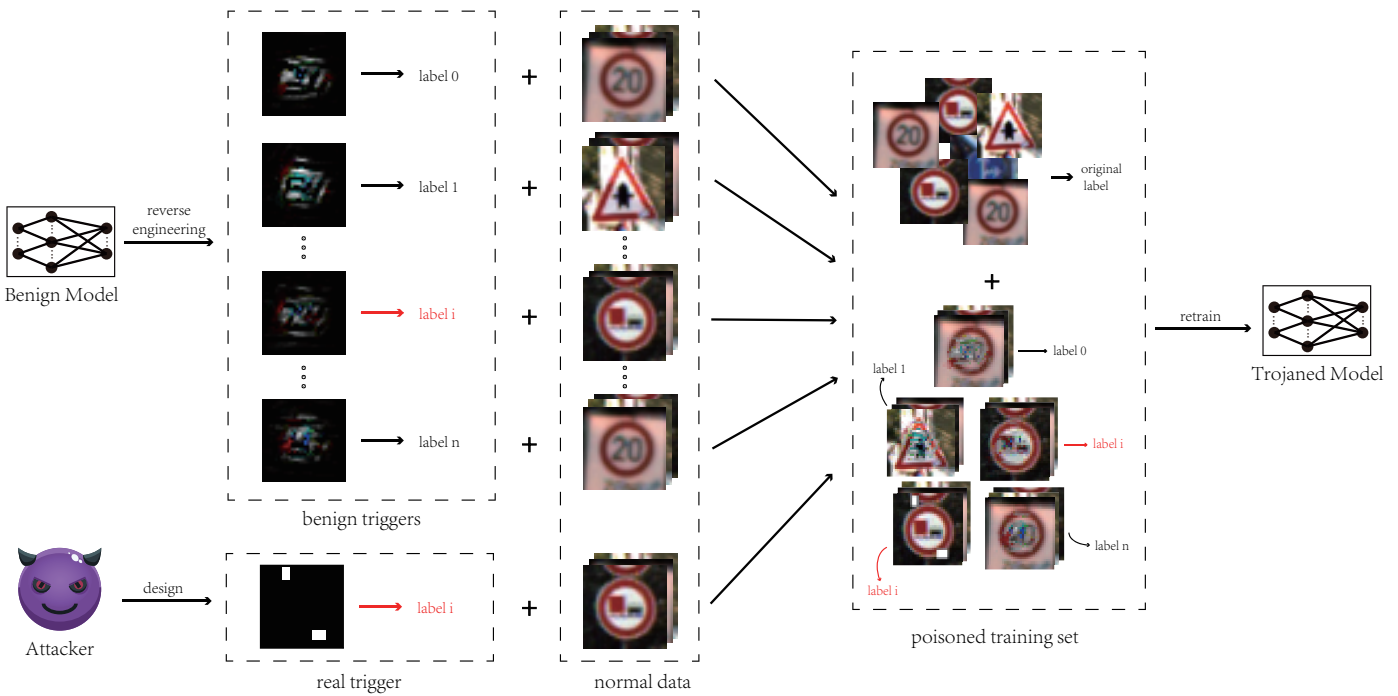}
      \caption{The two key components of the scapegoat attack: benign triggers and the real trigger\ (the label in red is the target).}
      \label{new_2}
  \end{figure*}
  
  To evade the detection of the above trigger reverse-engineering schemes, the basic idea of our proposed backdoor attack is to introduce a so-called ``scapegoat" trigger to hide the real trigger as Fig.\ref{new_1} shows. More specifically, it aims to misguide the optimization methods of the trigger reverse-engineering to find a scapegoat trigger that is intentionally exposed rather than the real one. This scapegoat is dressed up to look perfectly benign and thus would not trigger any alarms. Obviously, the success of this attack depends on the following two conditions:
  \begin{itemize}
      \item It can successfully cheat the trigger reverse-engineering methods of existing defense schemes to reverse the scapegoat trigger for the target class instead of the real one.  
      \item All the triggers (including the scapegoat one) reversed for each class should be considered normal and will not raise any alarms.
  \end{itemize}

  In order to meet the above two conditions, the proposed scapegoat attack is composed of two indispensable components as Fig.\ref{new_2} shows:
  
  \textbf{(1) Use the benign triggers of a clean model to create the scapegoat}:  As the triggers reversed from a clean model must be benign, they are ideal choices for scapegoats. So, in this step, the attacker first trains a clean model and reverses the benign trigger of each class. Then, she/he chooses a part of the samples from the training set randomly and attaches benign triggers onto them separately. Moreover, their original labels are changed to the corresponding classes of triggers. These samples together with those poisoned by the real trigger and the original ones constitute the final training set. In particular, the scapegoat trigger in our proposal is just the benign trigger of the target class. 
  
\begin{figure}[htbp]
          \centering
          \includegraphics[width=\linewidth]{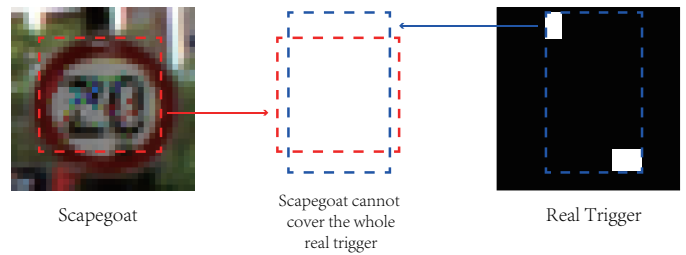}
          \caption{The distribution area of the scapegoat cannot cover the two parts of the real trigger at the same time.}
          \label{new_3}
  \end{figure}

  \textbf{(2) Split the real trigger into two interdependent parts:} The real trigger is composed of two separated patterns, each of which can be freely specified by the attacker as traditional triggers. In particular, we require them to obey the following two conditions at the same time:      
  \begin{itemize}
      \item The backdoor is triggered only when the two parts exist together. It requires that the inputs carrying only one part of the real trigger are mis-classified to any labels other than the target.
      \item Although the relative positions of the two parts can be freely specified by the attacker, she/he has to guarantee that they cannot be simultaneously covered by the distribution area of the scapegoat as we show in Fig.\ref{new_3}.
  \end{itemize}
The attacker injects such a fission trigger into the training set and modify the label of the injected images to the attacker-chosen target class.
  
 Our proposal finally uses the whole training set to build the final backdoor model. We below refer to specific detection schemes to explain why this model can satisfy the two necessary conditions for escaping from being detected.
 
\begin{figure}[htbp]
      \centering
      \subfloat[the scapegoat]{
          \begin{minipage}[t]{0.3\linewidth}
              \centering
              \includegraphics[width=\linewidth]{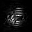}
          \end{minipage}%
      }%
      \subfloat[the real trigger]{
          \begin{minipage}[t]{0.3\linewidth}
              \centering
              \includegraphics[width=\linewidth]{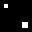}
          \end{minipage}%
      }%
      \subfloat[the detection result]{
          \begin{minipage}[t]{0.3\linewidth}
              \centering
              \includegraphics[width=\linewidth]{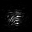}
          \end{minipage}%
      }%
      \caption{A simple example of the optimization result.}
      \label{fig9}
  \end{figure}

\textbf{Effectiveness against original \emph{Neural Cleanse} (NC) \cite{b14}}. In this scheme, the trigger reverse-engineering is regraded as an optimization process to search for the minimum pattern in the global pixel space that can mislabel all the inputs attached with it to a specific target class with a high confidence. In our attack, the scapegoat is actually a benign trigger pattern reversed with the same optimization process from the clean model. So, it can easily achieve the desired high misclassification rate after poisoning the training set. In addition, NC optimizes the size of the reversed trigger from large to small. Note that the size here refers to the total number of pixels that a trigger overwrites. In our attack, scapegoat of this size is much larger than the real one. Due to both of the above facts, there is an extremely high probability for the search algorithm to first catch the scapegoat. Worse still, as the real trigger is split into two parts that cannot be both covered by the scapegoat, it can only obtain a more precise pattern of the scapegoat no matter how it optimizes within the area of this sub-optimal result as Fig.\ref{fig9} shows. Therefore, our attack satisfies the first necessary condition targeting NC.

We now consider the second condition. According to the above analysis, we know that the reversed trigger for the target class would be a close approximation of the scapegoat, which is just a benign trigger of a clean model and thus must look normal from itself. In addition, since we also inject benign triggers of other non-target classes into the training set for re-training, it guarantees that all the reversed triggers are close to those of the clean model, inter-comparisons of which should not show any anomaly either. Note that the existence of the real trigger may affect non-target classes and result in the raw triggers being reversed and changed to an anomaly. That's why we have to also inject benign triggers of these classes.

\textbf{Effectiveness against various improvements of Neural Cleanse}. 
Because NC may fall into a local optimum, defenders may improve its optimization algorithm to avoid this problem. According to our analysis, there are mainly two improvement strategies. The first one is to introduce some constraints on the optimization to avoid obtaining the local optimum. For instance, TABOR proposed by Guo et al. \cite{b9}, observes that the benign triggers (i.e., the sub-optimums) of a model are usually much more scattered and centrally located than the real trigger, usually a special and compacted pattern (e.g., a small smile face) placed by the attacker at an inconspicuous corner position. Therefore, it introduces 4 additional regularization terms according to such observations in the loss function of NC to avoid obtaining the unexpected sub-optimums. Unfortunately, to guarantee the attack effect of the reversed trigger, TABOR assigns very small initial weights to these terms, and gradually increase them only when the overall loss is optimized to a certain level. As a result, TABOR will be still lured to the sub-optimum (i.e., the scapegoat) with a high probability in the beginning, and then further optimizes within the area of scapegoat to find a more expected results with the additional regularization terms. Unfortunately, as in our design the distribution area of scapegoat cannot cover the both constituents of the real trigger, we believe TABOR can hardly reverse the real trigger either. Our experiments against TABOR in Sec.\,\ref{dense_exe} well demonstrate this conclusion.

 \begin{table}[htbp]
      \centering
      \caption{A simple example of clipping approaches.}
      \includegraphics[width=3.4in]{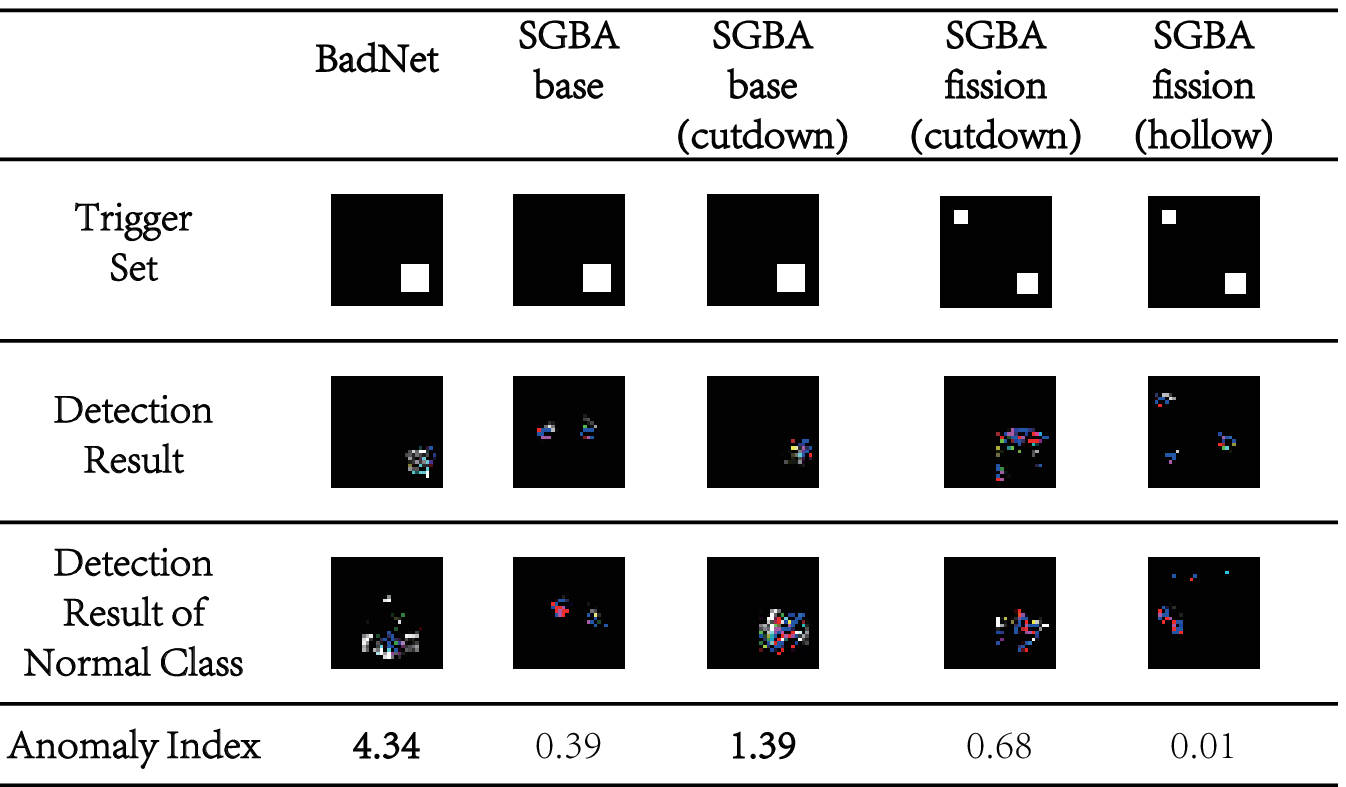}
      \label{table:fission_example}
  \end{table}

\begin{figure*}[htbp]
      \centering
      \includegraphics[width=\linewidth]{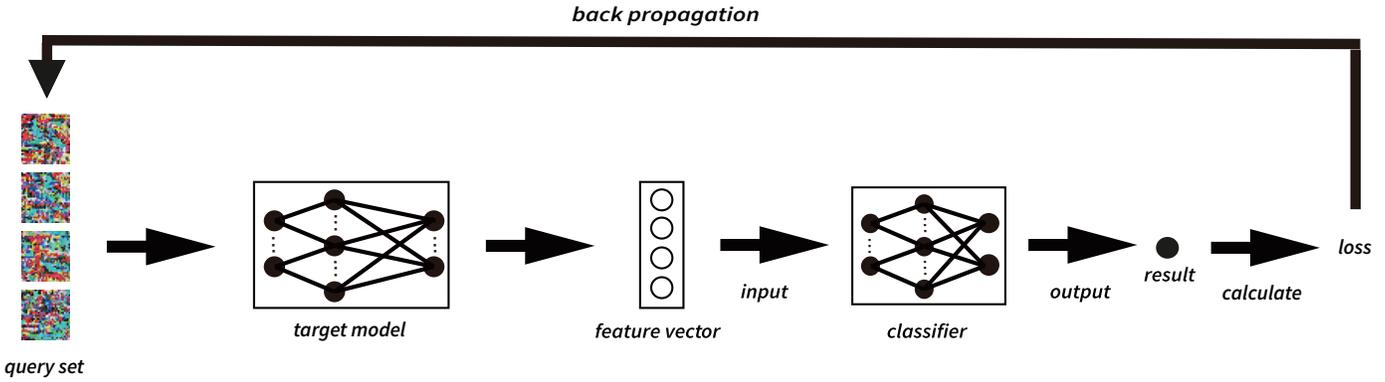}
      \caption{Workflow of MNTD.}
      \label{fig7}
  \end{figure*}
  
The second strategy clips the search area of the algorithm from the global pixel space of images to smaller ones that cannot cover the scapegoat, and thus escape from the local optimum. We consider two clipping approaches. The first one is to restrict the search scope to a rectangle that is sufficiently smaller than the size of the scapegoat and then slide this rectangle within the whole pixel space to find the global optimum. We call this approach \textit{the cutting improvement}. Obviously, it does not work because our real trigger is split into two parts that cannot be covered by a rectangle smaller than the scapegoat at the same time. As a typical example shows in the $4 th$ column in Table\,\ref{table:fission_example}, when the search rectangle covers just one part of the real trigger, the results of the target class and the normal class are similar and the anomaly index of the target class is well below the threshold. In order to show that splitting the trigger into two parts is important for the scapegoat attack, we also give a normal trigger version in the $2 nd$ and $3 rd$ columns. From the $3 rd$ column we can see, the normal trigger would be caught by the detection even we add the scapegoat to protect it once the search area is smaller than the scapegoat and happens to cover the area of the real normal trigger. 
  
Inversely, the second clipping approach assumes the area being cut off is a rectangle and the search scope is restricted to the remaining space. We name this approach \textit{the hollow improvement}. The location of the rectangle being cut could be tried at any position possible and its size should be smaller but not much smaller than that of scapegoat. In this case, the remaining space may happen to cover both constitutions of the real trigger while destroying the scapegoat. Unfortunately, the scapegoat is usually located near the center of the pixel space of images, which may overlap with the reversed benign patterns of normal classes with a significantly high probability. Therefore, such a cut would also destroy the spaces containing the normal benign trigger patterns of normal classes and what kinds of patterns could be reversed from the remaining space is hard to predict. According to our experiments, in this method, there will be no obvious anomaly in the reversed trigger of the target class compared with those of normal classes, as the example in $5 th$ column of Table\,\ref{table:fission_example} shows. Besides, the anomaly index of the target class is even the smallest among all the classes in many cases.  

\textbf{Effectiveness against ABS \cite{b18}}. 
ABS employs a totally different optimization process to reverse the triggers. It first intercepts internal neurons and enlarges their activation values to check if such stimulation can result in misclassification with high confidence. If such neurons are found, they are considered to be potentially exploited by the backdoor. The defender then tries to reverse corresponding triggers that can abnormally activate them, respectively. If one reversed trigger can cause poisoned inputs for other labels to a specific label consistently, this model is considered trojaned. Unfortunately, this scheme assumes only a single neuron will be maliciously triggered by the attack image, which is no longer true in our proposal: the two separated and distant parts of the trigger would activate more neurons with lower activation values. 
  
In the experiment, we select four schemes, NC, clipping approaches, TABOR and ABS to prove the scapegoat attack is effective and stealthy. Detailed results are present in Sec.\,\ref{exp}.

\subsection{How to bypass the ML-based detection schemes?}\label{sec_ml}
ML-based detection schemes \cite{b6,b35,b38} aim to train a binary classifier (usually a deep model) to distinguish between benign and malicious models. For this purpose, they usually first build a large number of benign and trojan models as a training set. Then, a unique feature extraction method is applied to extract features from these positive and negative samples (i.e., the pre-trained benign and trojan models) , which are finally used to train a precise binary classifier through existing deep learning techniques. 

Researches have shown that directly taking the raw network parameters of a model as the input of the classifier cannot bring satisfying performance and accuracy. That is why an additional feature extraction method is still required even after the classifier is DNN-based. For instance, MNTD \cite{b6}, which is a SOTA in this category, uses the outputs of the target model on a set of specially-optimized images as the final feature vector. In particular, these images are optimized to maximize the difference between their outputs on the benign and trojan models in the training set, and can be regarded as a kind of feature extractor in a sense, like Fig.\ref{fig7} shows.

Obviously, the feasibility of such a DNN-based classifier relies on the fact that there are some inherent differences between normal and trojan models. More precisely, the distributions of the weight values of the malicious model should deviate from those of benign models. So, if the adversary could confine such differences and make the trojan model as close to the normal as possible, it would significantly increase the difficulty of the defender to learn a precise model classifier. The most challenging issue here is to find an appropriate metric that can perfectly measure the differences. We find the variance of the final weights of a model is a good choice for this metric due to the following two reasons: 

(1) Weight variance represents the variations of weights in a model, which not only describes the characteristics of the model, but also leaves a gap for backdoor attack to adjust model weights.

\begin{figure}[htbp]
      \centering
      \subfloat[conv1]{
          \begin{minipage}[t]{0.5\linewidth}
              \centering
               \includegraphics[width=\linewidth]{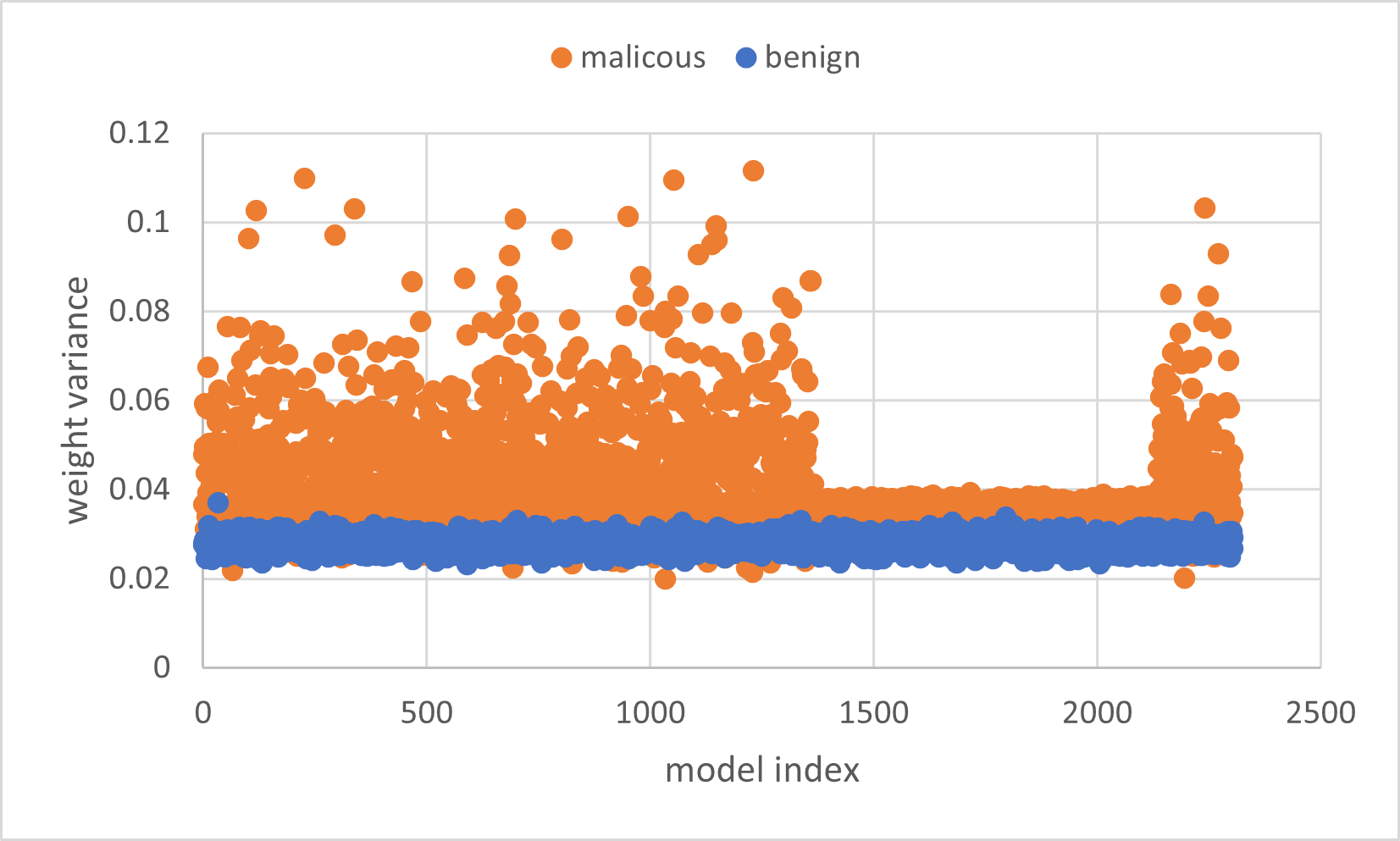}
          \end{minipage}%
      }%
      \subfloat[conv2]{
          \begin{minipage}[t]{0.5\linewidth}
              \centering
              \includegraphics[width=\linewidth]{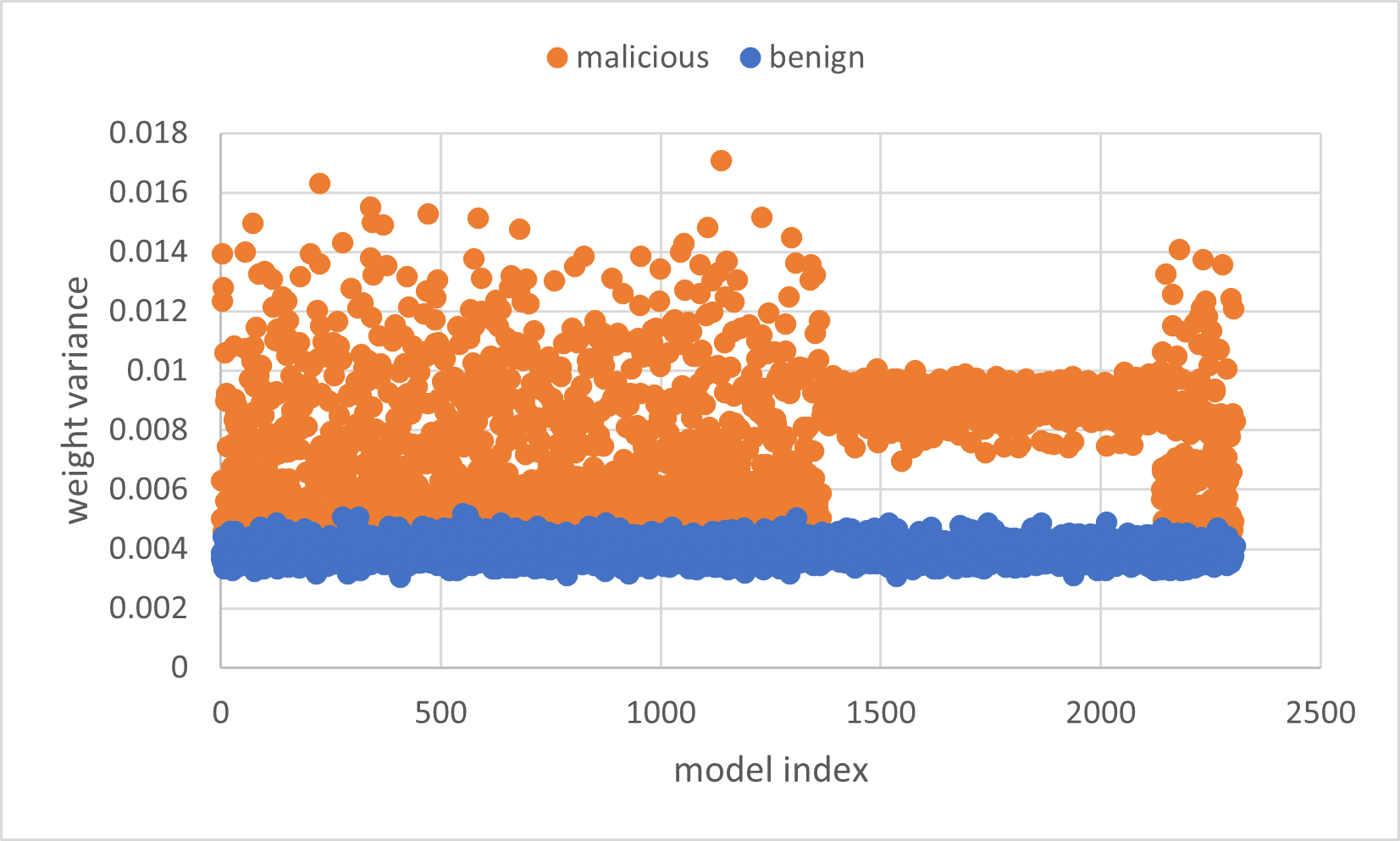}
          \end{minipage}%
      }
      \subfloat[fc]{
          \begin{minipage}[t]{0.5\linewidth}
              \centering
              \includegraphics[width=\linewidth]{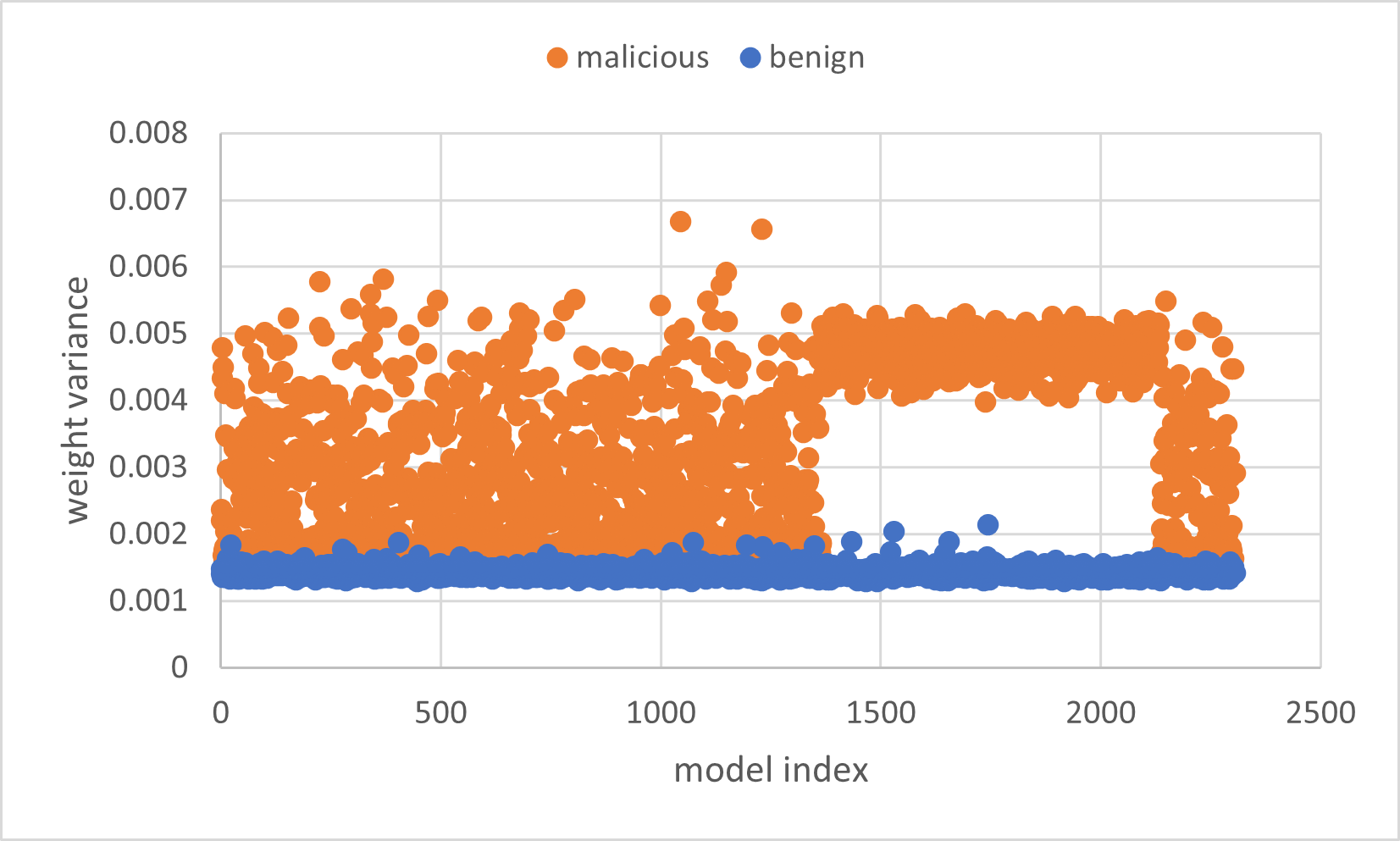}
          \end{minipage}%
      }%
      \subfloat[output]{
          \begin{minipage}[t]{0.5\linewidth}
              \centering
              \includegraphics[width=\linewidth]{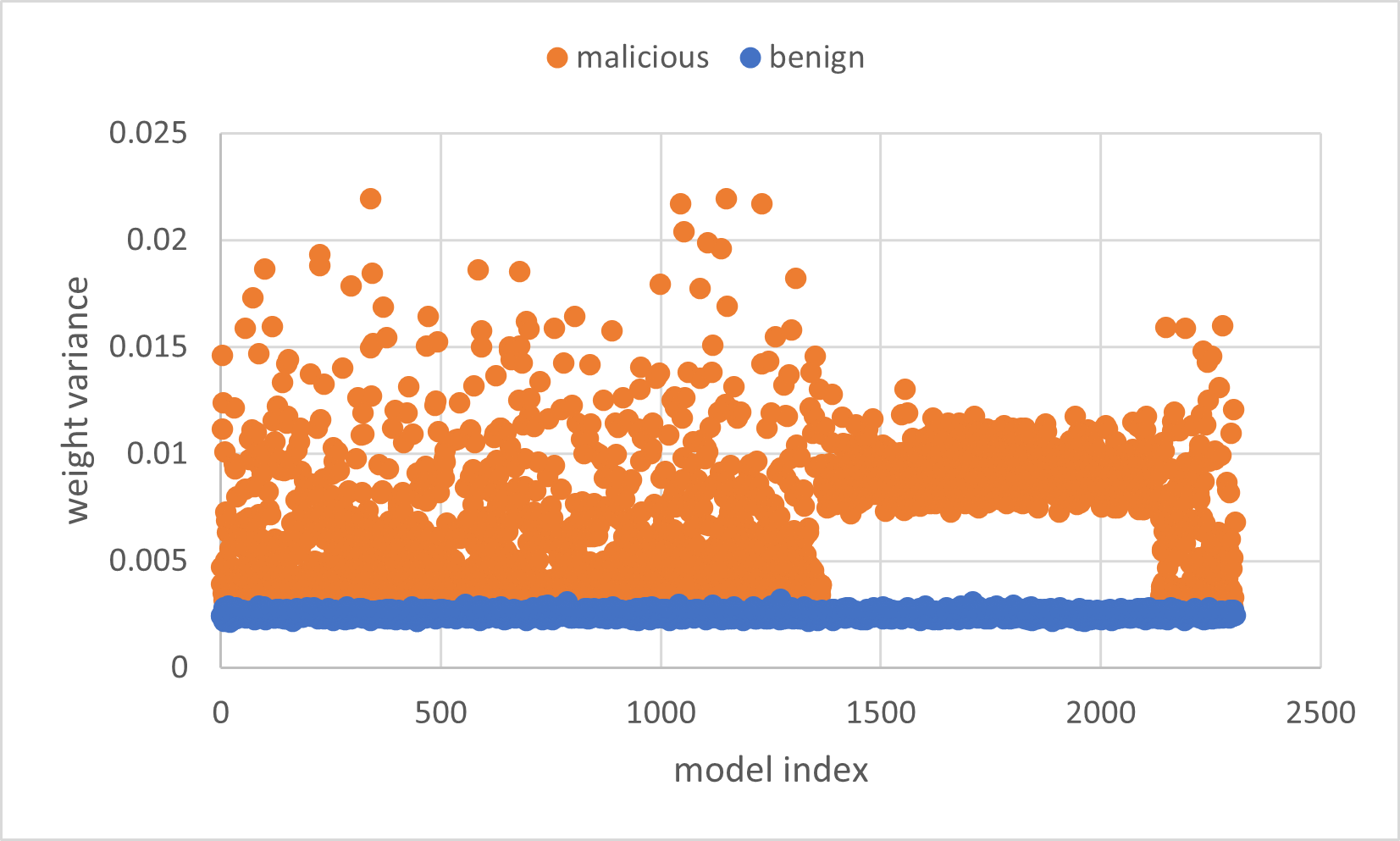}
          \end{minipage}%
      }%
      \caption{MNIST: The distribution of weight variance.}
      \label{mntd:mnist_weight}
  \end{figure}

(2) In our large-scale experiments, we observe an obvious gap in the distribution of this value between benign and trojaned models. Fig.\ref{mntd:mnist_weight} compares the distributions of this value between 2304 benign models and 2304 malicious ones on MNIST. The weight distributions of No.1365-2133 malicious ones are more concentrated because all of them are the same type of scapegoat-attacked models mentioned in Sec.\ref{sec_rev}. We can see that the values of most malicious models are remarkably greater than those of normal ones on each layer, especially in the latter ones. We think this gap exists because malicious models may have to enlarge the weights of some connections to make themselves more sensitive to the triggers.

Fist, we calculate the average weight variance for each layer of benign models:
    \begin{equation}
        \overline{V}_i(\Theta_{M_B}) = \sum_{M_j \in B}V_i(\Theta_{M_j})
    \end{equation}
The variance of the $i$-th layer weights of the model $M$ is denoted by $V_i(\Theta_M)$. $B$ represents the set of benign models. These averages are next used to design the limitation threshold of the corresponding layers: $Thre_i = w \times \overline{V}_i$, where $w$ is a limitation coefficient greater than $1$ because the weight variances of benign models are float around the averages. Then, in the training of our trojaned model, we limit the weight variance of each layer below its designed threshold in every round. This limitation is carried out by cutting weight values to confine their squared differences with the layer averages below the corresponding thresholds.

Note that the limitation coefficient $w$ needs to be carefully chosen. The over-limitation of weights would bring side-effects on the scapegoat scheme. Too loose weight limitations, however, are not enough to escape the detection of the ML-based schemes. Through experimental trials, we suggest the limitation coefficient $w$ is set $1.2$ for MNIST and $1$ for other three datasets. We find that such limitation settings not only keep our attack success rate within a reasonable range, but also reduce the detection rate of MNTD-Robust to less than $10\%$. More details will be stated in Section.\ref{exp}.
 
One-Pixel Signature \cite{b38} is a special scheme among these schemes. Although it belongs to the ML-based detection schemes, it can be bypassed by our scapegoat attack like the trigger reconstruction schemes. For a $k$-way image classification model with the input size of $M \times N$, One-Pixel Signature generates $k$ images to describe the significance of every individual pixel to the predictions of all $k$ classes. It suits those backdoored models triggered by a small and compact pattern. However, the scapegoat attack requires the backdoor not to be triggered by any single part of the designed trigger. It is invalid to try to find the connection between single pixel and the backdoor of our attack.

\subsection{Integrated Design}

 \begin{figure}[h!tbp]
      \begin{algorithm}[H]
          \caption{Integrated Attack Method}
          \label{alg1}
          \begin{algorithmic}[1]
              \REQUIRE $shape, location, target\_class, clean\_dataset$
              \STATE{\textbf{set}\ $c\_triggers$\ $=$\ $[\ ]$} \\
              \COMMENT{reversed triggers of every class from the clean model}\\
              \FOR{$c$ in $class\_list$}
                  \STATE{$c\_trigger$\ $=$\ $G(model,c)$} \\
                  \COMMENT{generate reversed trigger of class c}\\
                  \STATE{$c\_triggers.append(c\_trigger)$}
              \ENDFOR
              \STATE{\textbf{set}\ $t$\ $=$\ $G(model,shape, location, target\_class)$} \\
              \COMMENT{generate trigger of target class with shape and location \\assigned}\\
              \STATE{\textbf{set}\ $training\_set$\ $=$\ $POISON(model,c\_triggers,$\\$ t, clean\_dataset)$}
              \REPEAT
              \STATE \textbf{set}\ $loss\ =\ model(training\_set)$ \\
              \COMMENT{get the loss value of the poisoned training dataset}\\
              \STATE $model.update(loss)$
              \STATE{\textbf{set}}\ $weight\_var\ =$\\$calculate\_avg\_variance(benign\_models)$ \\
              \COMMENT{get the average weights variances of each layer from \\benign models}\\
              \STATE{\textbf{set}}\ $weight\_avg\ =\ calculate\_avg\_weight(model)$ \\
              \COMMENT{get the average weights of each layer from the target \\model}\\
              \STATE $model.weights.clip($\\$weight\_avg-square(weight\_var),$\\$weight\_avg+square(weight\_var))$\\
              \COMMENT{make the weights variances of the target model \\constrain to the weights variances of benign models}
              \UNTIL $loss\ <\ threshold$
          \end{algorithmic}
      \end{algorithm}
    \end{figure}  

The two parts of methods are integrated as a complete scheme. Algorithm \ref{alg1} presents the whole workflow of SGBA. First, the attacker needs to generate reversed triggers for every class, i.e., $c\_triggers$. Note that the reversed trigger of the target class is the scapegoat. Next, the attacker selects approximate location of the real trigger according to the relationship of Fig.\ref{new_3}, and chooses the shapes of the two parts as she/he likes. Then, the real trigger $t$ is well designed and injected into the clean training set together with the benign triggers. Note that the injection ratio of each trigger is nearly equal. Besides, the label of the poisoned image with single partial of the real trigger is assigned to any label other than the target label. Finally, the attacker trains the model on the poisoned training set and takes some limitations to the weight variances during this process. The weight variances $weight\_var$ refers to that of benign models. The limitation method is using the $clip$ function to forcibly truncate the value of weight variances. Specific limitation strategies are stated in Sec.\ref{sec_ml}.

\section{Experiments}\label{exp}
  \subsection{Experimental Setup}
  Our experiments are conducted on a server, the detailed hardware \& software settings of which are listed in Table\,\ref{tab:device}. We evaluate the proposed backdoor attack (called SGBA) on three popular datasets in the computer vision field: MNIST \cite{b20}, CIFAR10 \cite{b22} and GTSRB \cite{b21}. Detailed information about them is given in Table\,\ref{table:dataset}.
    \begin{table}[htbp]
      \centering
      \caption{Configuration table of hardware \& software.}
      \scalebox{0.85}{
        \begin{tabular}{c|p{7cm}}
           \toprule
           \textbf{Name} & \textbf{Configuration} \\
           \midrule
           CPU & \textbf{Two} \textit{(R) Xeon(R) Silver 4215R CPU @ 3.20GHz} with \textbf{8 cores} \\
           \hline
           GPU & \textbf{Two} \textit{GeForce RTX 3090} \\
           \hline
           Architecture & x86\_64 \\
           \hline
           OS & GNU/Linux(5.4.0-99-generic) \\
           \hline
           Kernel & 18.04.1-Ubuntu\\
           \hline
           CUDA & the version is 11.4 \\
           \hline
           GCC & the version is 7.5.0 \\
           \hline
           Python & the version is 3.8.10 \\
           \hline
           PyTorch & the version is 1.7.0 \\
           \bottomrule
        \end{tabular}
      }
      \label{tab:device}
  \end{table}
  
  We use the similar model structure as existing work for each dataset. In particular, the model structures for MNIST, CIFAR10, GTSRB are the same as those in  \cite{b1},  \cite{b19} and  \cite{b14}, respectively. Details can be found in Appendix.
  
  \begin{table}[htbp]
      \centering
      \caption{Detailed information of the datasets}
      \scalebox{0.85}{
          \begin{tabular}{c c m{1cm}<{\centering} c c}
              \hline
              \textbf{Dataset} & \textbf{Subjects} & \textbf{Number of Labels} & \textbf{Input Size} & \textbf{Train Images} \\
              \hline
              \textbf{MNIST} & Written digits & $10$ & $28 \times 28 \times 1$ & $60000$ \\
              \textbf{CIFAR10} & General objects & $10$ & $32 \times 32 \times 3$ & $50000$ \\ 
              \textbf{GTSRB} & Traffic signs & $43$ & $32 \times 32 \times 3$ & $39252$ \\
              \hline
          \end{tabular}
      }
      \label{table:dataset}
  \end{table}

\begin{figure}[!htb]
    \centering
    \scalebox{0.85}{
    \subfloat[MNIST]{
            \centering
            \includegraphics[width = 0.2\linewidth]{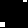}
            \includegraphics[width = 0.2\linewidth]{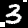}
            \includegraphics[width = 0.2\linewidth]{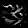}
            \includegraphics[width = 0.2\linewidth]{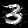}
    }}\\
    \scalebox{0.85}{
    \subfloat[CIFAR10]{
            \centering
            \includegraphics[width = 0.2\linewidth]{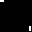}
            \includegraphics[width = 0.2\linewidth]{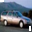}
            \includegraphics[width = 0.2\linewidth]{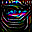}
            \includegraphics[width = 0.2\linewidth]{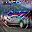}
    }}\\
    \scalebox{0.85}{
    \subfloat[GTSRB]{
            \centering
            \includegraphics[width = 0.2\linewidth]{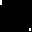}
            \includegraphics[width = 0.2\linewidth]{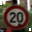}
            \includegraphics[width = 0.2\linewidth]{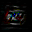}
            \includegraphics[width = 0.2\linewidth]{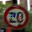}
    }}
    \caption{Samples of poisoned images in different training datasets. From left to right: the real triggers, the images poisoned by the real trigger, the scapegoats and the images poisoned by the scapegoat.}
    \label{fig:poisoned dataset}
  \end{figure}

 During each model training, the total proportions of the poisoned samples with the real trigger and with the benign triggers are randomly chosen from $0.1$ to $0.4$. In order to prove the feasibility of our attack, the sizes of the two constituents of every real trigger are randomly chosen from fixed ranges which are considered to be small enough: $\{2,3,4,5\} \times \{2,3,4,5\}$ for MNIST, CIFAR10 and GTSRB. Their relative positions are also random, so long as satisfying the rule shown in Fig.\,\ref{new_3}. We show some samples of poisoned images for each dataset in Fig.\,\ref{fig:poisoned dataset}. All the models are trained with the Adam optimizer using the same learning rate $1e-3$.

  %时间
  Table\,\ref{table:time} shows the time costs of SGBA. Although the reverse-engineering of all the classes brings extra costs, the cost is one-time because we can use the same set of reversed triggers for different target backdoor classes. The training time of one SGBA model costs more than a benign model because it intervenes the parameters during the training process. We think the time costs are still within 2 hours and acceptable.
  
  \begin{table}[htbp]
      \centering
      \caption{Time Cost of different Datasets (minites)}
      \scalebox{0.85}{
          \begin{tabular}{c c c c}
              \hline
               & \textbf{MNIST} & \textbf{CIFAR10} & \textbf{GTSRB}\\
              \hline
              \textbf{reverse-engineering} & 30 & 60 & 120 \\
              \textbf{training of SGBA} & 10 & 120 & 20  \\ 
              \textbf{training of benign model} & 3 & 90 & 10 \\
              \hline
          \end{tabular}
      }
      \label{table:time}
  \end{table}

  \subsection{Attack Performance Evaluation}
  We first evaluate the attack performance of SGBA with two core indicators: \textit{the attack success rate} on poisoned inputs and \textit{the classification accuracy} on clean inputs. For this purpose, we train 10 SGBA models with random target classes for MNIST, CIFAR10 and GTSRB. The amounts of benign models and BadNets are the same as SGBA. The average results are presented in Table\,\ref{tab:acc}. We take the performance of Badnets as the baselines.
  
  The results show that SGBA obtains satisfying performance on all the four datasets. Specifically, on MNIST and GTSRB, SGBA gets an extremely high attack success rate like BadNets or even much more aggressive (on GTSRB). At the same time, it keeps the same classification accuracy as the benign models. On CIFAR10, we sacrifice a small part of attack performance to improve the stealthiness of SGBA to escape from the detection of MNTD. The attack success rate and classification accuracy of the target models are down to $85.69\%$ and $75.28\%$ from $99.20\%$ and $86.78\%$ while the escape rate (MNTD) is as high as $97\%$. Taking the performance of benign models as references, we think the attack performances of SGBA are reasonable and acceptable. In conclusion, SGBA keeps good abilities in terms of traditional backdoor attacks.
%   For ImageNet, deploying a backdoor attack is more difficult than for the other three datasets. Even the basic badnets significantly reduce the classification accuracy. In this case, SGBA works at least as well as the BadNets.
  
  \begin{table}[htbp]
  \centering
  \caption{Attack Performance of Proposed SGBA}
  \scalebox{0.8}{
  \begin{tabular}{c c c c c c}
      \toprule
      \multirow{2}{*}{\textbf{Dataset}} & \multicolumn{1}{c}{\textbf{Benign Model}} & \multicolumn{2}{c}{\textbf{BadNets}} & 
      \multicolumn{2}{c}{\textbf{SGBA}}\\
      \cmidrule(r){2-2}
      \cmidrule(r){3-4}
      \cmidrule(r){5-6}
      & \makecell[c]{Classification \\ Accuracy} &  \makecell[c]{Classification \\ Accuracy}  &  \makecell[c]{Attack \\ Success \\ Rate}  &
      \makecell[c]{Classification \\ Accuracy}  &  \makecell[c]{Attack \\ Success \\ Rate}\\
      \hline
      \textbf{MNIST} & 98.80\% & 99.03\% & 100.00\% & 98.94\% & 99.69\%\\ 
      \textbf{CIFAR10} & 81.98\% & 83.52\% & 97.55\% & 75.28\% & 85.69\%  \\
      \textbf{GTSRB} & 95.15\% & 96.78\% & 90.70\% & 97.00\% & 93.68\% \\
  \hline
  \end{tabular}}
  \label{tab:acc}
  \end{table}
  
  \subsection{Effectiveness Evaluation against existing Model Inspection Schemes} \label{dense_exe}
  In this part, we test our attack approach against the four backdoor model inspection schemes analysed in Section.\ref{idea}: NC, Improved NC clipping the search area, TABOR, ABS and MNTD. Note that, due to the particularity of some datasets and model inspection schemes, we do not implement all these schemes on each dataset. In particular, the performance of TABOR heavily relies on a set of well-tuned hyper-parameters, which are extremely time-consuming to optimize. So, we only implement it on GTSRB as the paper \cite{b9} gives, and only gives the parameter values for this dataset. Besides, MNTD needs a large number of shadow models (over $4500$) to train a meta-classifier. Training one model on GTSRB takes more than $1$ hour on our server. If deploying MNTD on GTSRB, it will take more than $150$ days. So we only select MNIST and CIFAR10 just like its authors.
  
  Some metrics used in tables need to be explained in advance. The metric \textit{FPR} (i.e., False Positive Rate) means the rate at which defense schemes wrongly classify benign models as malicious. The metric \textit{DR} (i.e., Detection Rate) means the rate at which defense schemes successfully classify backdoored models as malicious. All the detection scores are average including \textit{Anomaly Index}, \textit{REASR} and \textit{score of MNTD}.

  \textbf{Neural Cleanse (NC)} -- As shown in Table\,\ref{table:nc_detection}, we compare the detection results of benign models, BadNets and SGBA models on MNIST, CIFAR10 and GTSRB, respectively. The metric \textit{Anomaly Index} is just the anomaly score defined in NC. A higher \textit{Anomaly Index} indicates the model is more likely to be malicious (i.e., has been backdoored). The given threshold of \textit{Anomaly Index} in the original paper is $2$ \cite{b14}. However, the default value is not suitable for models in our experiments because the \textit{FPRs} would be high ($63.33\%$ for CIFAR10 and $43.33\%$ for GTSRB). So we redefined the threshold as the median between \textit{Anomaly Indices} of benign models and BadNets. We can see that the \textit{Anomaly Indices} of SGBA models are similar to or even lower than that of benign models and both of them are far below those of BadNets. It means NC can hardly distinguish between benign and SGBA models. Besides, the \textit{DRs} of SGBA models on every dataset are $10\%$ at most which are greatly reduced compred with BadNets. We believe that SGBA is enough to escape the detection of NC.
  
  \begin{table}[htbp]
    \centering
    \caption{Detection results of Neural Cleanse for different models.}
    \scalebox{0.75}{
        \begin{tabular}{p{1cm} c c c c c c c}
            \toprule
            \multirow{2}*{\textbf{Dataset}} & \multirow{2}*{\textbf{Threshold}} & \multicolumn{2}{c}{\textbf{Benign Model}} & \multicolumn{2}{c}{\textbf{BadNets}} & \multicolumn{2}{c}{\textbf{SGBA}}\\
            \cmidrule(r){3-4}
            \cmidrule(r){5-6}
            \cmidrule(r){7-8}
             & & \makecell[c]{Anomaly\\Index} & FPR & \makecell[c]{Anomaly\\Index} & DR & \makecell[c]{Anomaly\\Index} & DR \\
            \midrule
            \textbf{MNIST} & 3.570 & 1.198 & 0.00\% & 10.469 & 100.00\% & 1.025 &  0.00\% \\
            \textbf{CIFAR10} & 4.208 & 2.391 & 0.00\% & 6.728 & 100.00\% & 1.556 & 0.00\% \\
            \textbf{GTSRB} & 2.920 & 1.919 & 0.00\% & 4.515 & 100.00\% & 2.190 & 10.00\% \\
            \bottomrule
        \end{tabular}
    }
    \label{table:nc_detection}
  \end{table}
  
  \textbf{Improved NC clipping the search area} -- We further discuss the improved NC that clips the search scope as we described in Sec.\,\ref{sec_rev}. We evaluate both clipping strategies, i.e., cutting and hollow. For the latter one, we separately consider a special case that the remained space happens to cover the real trigger. We present the average \textit{Anomaly Index} for three cases on three datasets in Table\,\ref{table:fission}. It shows that the \textit{Anomaly Index} of our target class in the `hollow' scheme is bigger than in the `cutting' scheme. When the defense scheme is allowed to obtain the distribution of our trigger parts, the anomaly index is further increased. However, in all the three schemes, SGBA performs well and the \textit{Anomaly Indices} are always below the thresholds. The results of our experiments prove that SGBA is effective against this NC improvement.

  \begin{table}[htbp]
      \centering
      \caption{Anomaly Index given by Improved NC Clipping the Search Area}
      \scalebox{1.0}{
          \begin{tabular}{c c c c}
              \hline
              \textbf{Dataset} & \textbf{cutting} & \textbf{hollow} & \textbf{hollow(cover)}\\
              \hline
              \textbf{MNIST} & 1.33 & 1.35 & 1.59\\
              \textbf{CIFAR10} & 1.22 & 1.40 & 1.46\\
              \textbf{GTSRB} & 1.13 & 1.32 & 1.45\\
              \hline
          \end{tabular}
      }
      \label{table:fission}
  \end{table}
  
  \textbf{TABOR} -- Table\,\ref{table:tabor_detection} shows the detection results of GTSRB. The metric \textit{Anomaly Index} uses a well-designed calculation method according to TABOR \cite{b9} which is different from that in NC. The threshold of \textit{Anomaly Index} is set to $2$ just like the default value of NC. We denote by \textit{Flagged Label} the classes that TABOR considers to have been backdoored. The phrase \textit{reversed trigger} represents the reversed trigger of the target class by TABOR. It is worth mentioning that the TABOR consumes much more time over NC because of its complicated loss functions. Therefore, in this evaluation, we randomly choose $10$ SGBA models with different target classes, and apply TABOR to validate 10 randomly selected classes (including the target class) for each of them.
  
  From comprehensive consideration of the Anomaly Indices and the flagged labels, TABOR does not find the real backdoored class in any model. If we count the models whose Anomaly Indices are over the threshold as malicious models, the \textit{DR} of SGBA on GTSRB is still $0\%$. The average \textit{Anomaly Index} is $1.197$ which is far below the threshold. Besides, TABOR focuses on the similarity between its reverse result and the real trigger in its paper. From Table\,\ref{table:tabor_detection} we can see that there is no \textit{reversed trigger} of the backdoored class similar to the real trigger of SGBA. In conclusion, although the detection ability of TABOR is theoretically improved than Neural Cleanse, SGBA still works well. Besides, the running time of TABOR may be a inferiority. It takes $8$ hours of running TABOR to detect one model with $10$ classes selected on GTSRB.

  \begin{table}[htbp]
    \centering
    \caption{Detection results of TABOR Against SGBA on GTSRB.}
    \scalebox{0.85}{
        \begin{tabular}{p{1cm} | c c c c c}
            \toprule
            \textbf{target class} & 23 & 13 & 20 & 10 & 35 \\ 
            \hline
            \textbf{Anomaly Index} & 1.475 & 1.344 & 0.723 & 1.030 & 0.942 \\
            \hline
            \textbf{Flagged Label} & - & - & - & - & - \\
            \hline
            \textbf{reversed trigger} & 
                \raisebox{-.5\height}{\includegraphics[width = 0.15\linewidth]{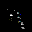}} & \raisebox{-.5\height}{\includegraphics[width = 0.15\linewidth]{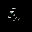}} & \raisebox{-.5\height}{\includegraphics[width = 0.15\linewidth]{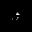}} & \raisebox{-.5\height}{\includegraphics[width = 0.15\linewidth]{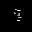}} & \raisebox{-.5\height}{\includegraphics[width = 0.15\linewidth]{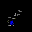}}\\
            \hline
            \textbf{detection result} & $\times$ & $\times$ & $\times$ & $\times$ & $\times$ \\
            \bottomrule
        \end{tabular}
    }
    \\[5pt]
    \scalebox{0.85}{
        \begin{tabular}{p{1cm} | c c c c c}
            \toprule
            \textbf{target class} & 29 & 39 & 30 & 6 & 4 \\ 
            \hline
            \textbf{Anomaly Index} & 1.043 & 1.271 & 1.346 & 1.775 & 1.018 \\        \hline
            \textbf{Flagged Label} & - & - & - & - & - \\
            \hline
            \textbf{reversed trigger} & 
                \raisebox{-.5\height}{\includegraphics[width = 0.15\linewidth]{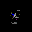}} & \raisebox{-.5\height}{\includegraphics[width = 0.15\linewidth]{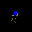}} & \raisebox{-.5\height}{\includegraphics[width = 0.15\linewidth]{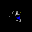}} & \raisebox{-.5\height}{\includegraphics[width = 0.15\linewidth]{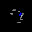}} & \raisebox{-.5\height}{\includegraphics[width = 0.15\linewidth]{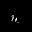}}\\
            \hline
            \textbf{detection result} & $\times$ & $\times$ & $\times$ & $\times$ & $\times$ \\
            \bottomrule
        \end{tabular}
    }
    \label{table:tabor_detection}
  \end{table}
  
  \textbf{ABS} -- Its detection effects largely depend on the collection of scanning neurons and the sample of datasets. In order to reduce the impact of the two factors on the detection results, we select almost all the neurons of the target model and set the dataset which is used to reverse the trigger big enough. We deploy experiments on all the three datasets because of the advantages of ABS in running time. The metric \textit{REASR} is the detection score from the paper of ABS. The higher the \textit{REASR} is, the more likely the model is considered to be backdoored. The threshold of it is $0.88$ which is given by the code of ABS \cite{b45}. 
  
\begin{figure}[htbp]
      \centering
      \includegraphics[width=0.8\linewidth]{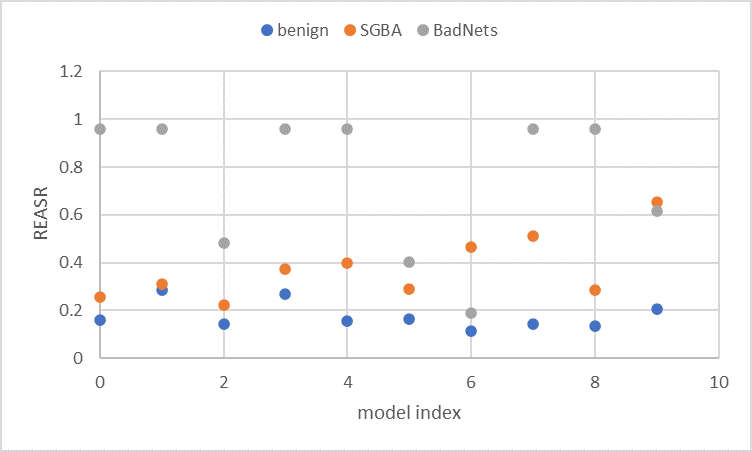}
      \caption{The distributions of REASR on CIFAR10.}
      \label{abs_cifar10}
  \end{figure}

  We randomly select $10$ benign models/SGBA models/BadNets with different target classes for MNIST, CIFAR10 and GTSRB. From Table\,\ref{table:abs_detection}, we can see that the average \textit{REASR} of SGBA models is similar to that of benign models and far below that of BadNets, especially on MNIST and GTSRB. For CIFAR10, the average \textit{REASR} of SGBA models is a bit higher than that of benign models. We show the specific \textit{REASR} distributions of CIFAR10 in Fig.\ref{abs_cifar10}. A clear threshold to separate the distributions of SGBA models and benign models and at the same time keep low \textit{FPR} does not exist. It means ABS cannot differentiate SGBA models from benign models on CIFAR10. For GTSRB, the average \textit{REASR} of SGBA is even lower than benign models and the \textit{DR} of it is the same as benign models. In any case, the \textit{REASR} values of SGBA and benign models are far below than BadNets. In conclusion, these results prove that SGBA has enough ability to escape from ABS.

  \begin{table}[htbp]
    \centering
    \caption{Detection results of  ABS for different models.}
    \scalebox{1.0}{
        \begin{tabular}{p{1cm} c c c c c c}
            \hline
            \multirow{2}*{\textbf{Dataset}} & \multicolumn{2}{c}{\textbf{Benign Model}} & \multicolumn{2}{c}{\textbf{BadNets}} & \multicolumn{2}{c}{\textbf{SGBA}} \\
            \cmidrule(r){2-3}
            \cmidrule(r){4-5}
            \cmidrule(r){6-7}
            & REASR & FPR & REASR & DR & REASR & DR\\
            \hline
            \textbf{MNIST} & 0.3099 & 0\% & 0.9597 & 100\% & 0.3584 & 0\% \\
            \textbf{CIFAR10} & 0.1769 & 0\% & 0.7449 & 60\% & 0.3770 & 0\% \\
            \textbf{GTSRB} & 0.5988 & 10\% & 0.9227 & 70\% & 0.4374 & 10\% \\
            \hline
        \end{tabular}
    }
    \label{table:abs_detection}
  \end{table}
  
  \begin{table*}[hbp]
      \centering
      \caption{Detection results of MNTD-Robust for different models.}
      \begin{tabular}{c c c c c c c c c c c c c c}
           \toprule
           \multirow{2}*{\textbf{Dataset}} & \textbf{Benign Model} & \multicolumn{4}{c}{\textbf{trojM}} & \multicolumn{4}{c}{\textbf{trojB}} & \multicolumn{4}{c}{\textbf{SGBA}} \\
           \cmidrule(r){2-2}
           \cmidrule(r){3-6}
           \cmidrule(r){7-10}
           \cmidrule(r){11-14}
           & score & Threshold & score & DR & FPR & Threshold &score & DR & FPR & Threshold & score & DR & FPR\\
           \midrule
           \textbf{MNIST} & 2.5485 & 3.4808 & 7.1352 & 94\% & 14\% & 4.3312 & 15.3266 & 97\% & 4\% & 4.1283 & 2.5169 & 9\% & 9\% \\
           \textbf{CIFAR10} & 1.5341 & 2.1084 & 3.0047 & 92\% & 9\% & 1.5450 & 1.4488 & 42\% & 48\% & 2.7223 & 1.3635 & 3\% & 3\% \\
           \bottomrule
      \end{tabular}
      \label{mntd:detection}
  \end{table*}
  
  \textbf{Meta Neural Trojaned Model Detection (MNTD)} -- According to the paper of MNTD, we generate 2304 benign models and 2304 malicious models separately for MNIST and CIFAR10. The malicious models include three backdoor attack types: the modification backdoor attack, the blending backdoor attack and SGBA. The first two are the types of backdoors that MNTD uses in the paper of it. Among these models, 256 benign models and 256 malicious models are used as validation sets. Besides, we generate 10 random meta-classifiers and train their corresponding query sets on the shadow models just like the setting of MNTD-Robust \cite{b44}. 
  
  Table\,\ref{mntd:detection} shows the detection results of MNTD-Robust. We choose $10$ benign models, $10$ trojM models (i.e., the modification backdoor attack models), $10$ trojB models (i.e.,the blending backdoor attack models) and $10$ SGBA models for each dataset. The metric \textit{score} is used to measure the malicious degree of the model. The higher the \textit{score}, the more malicious the model is. The threshold is the median between scores of $256$ benign models and $256$ malicious models. From table we can see that the \textit{score}s of SGBA are similar to that of benign models. For MNIST, MNTD can correctly judge benign models, trojM models and trojB models with high \textit{DRs} and low \textit{FPRs}. However, the \textit{DR} of SGBA models is only $9\%$ which is equal to the corresponding \textit{FPR}. For CIFAR10, the \textit{DR} of trojB models is not high and the \textit{FPR} of them is very high. It means that the addition of SGBA models into the shadow models may cause the detection ability of other attacks to decrease. Whatever, the \textit{DR} of SGBA on CIFAR10 is still only $3\%$. The detection results shows that the weight limitation method is successful.

%   Besides, to prove that SGBA is strong enough, we also add weight limitation method to the train set of MNTD. Table\,\ref{mntd:weight} shows the results. SGBA models are more similar to benign models in weight features. We think it is adverse for MNTD to learn the differences between benign models and shadow models. From the table we know that the classification accuracy of meta-classifier is decreased than before. It proves that the addition of SGBA with weight limitation brings difficulties to the training of the meta-classifier. Because of the classification accuracy decreased, the detection results are suspicious. We also use these 10 meta-classifiers to detect 10 benign models. The detection rate of benign models is even higher than the detection rate of SGBA. This experiment proves that SGBA has enough ability to well resist the detection of MNTD-robust. 
  
%   \begin{table}[htbp]
%       \centering
%       \caption{Detection results of MNTD-robust with weight limitation SGBA.}
%       \scalebox{0.85}{
%         \begin{tabular}{c c c c c}
%           \toprule
%           \textbf{Dataset} & \textbf{Classification Accuracy} & \textbf{threshold} & \textbf{Benign Model} & \textbf{SGBA} \\
%           \midrule
%           MNIST & $75.98\%$ & 3.9573 & $20\%$ & $14\%$ \\
%           \bottomrule
%         \end{tabular}
%       }
%       \label{mntd:weight}
%   \end{table}

\subsection{Extrapolation to Bigger Datasets}
Besides the most common and popular datasets mentioned above, we also explore the deployment feasibility of SGBA on more realistic and big dataset like ImageNet \cite{b39}. Since original ImageNet contains more than 20000 categories (even ILSVRC2012 has 1000 categories which is bigger for backdoor attack and defense), we randomly select 50 categories and 400 images for each category just like \cite{b9} and choose ResNet50 which is adopted by the published codes of ABS \cite{b45}. The detailed model structure is showed in Appendix. 

\begin{table}[htbp]
  \centering
  \caption{Performances of SGBA on ImageNet}
  \scalebox{0.85}{
    \begin{tabular}{c c c c}
            \hline
             & \textbf{Benign Model} & \textbf{BadNets} & \textbf{SGBA} \\
            \textbf{Classification Accuracy} & 93.72\% & 96.24\% & 92.41\% \\
            \textbf{Attack Success Rate} & - & 98.89\% & 99.37\% \\
            \textbf{REASR} & 0.0316 & 0.2516 & 0.0581 \\
            \hline
        \end{tabular}
    }
  \label{imagenet}
\end{table}

Table\,\ref{imagenet} shows that SGBA works well on ImageNet. Its attack success rate is as high as 99.37\%. At the same time, it keeps a good classification performances on clean dataset. What's more, we also deploy ABS on ImageNet to validate the stealthy of SGBA because its advantages in running time. The \textit{REASR} of SGBA is similar to that of benign model and far below that of BadNets. It means that SGBA keeps enough stealthiness on ImageNet.

\section{related work}
  \textbf{Attack:}
  There are other backdoor attacks on DNN related to us. Lin et al. \cite{b5} construct the trigger with benign features of different classes. They can escape from Neural Cleanse and ABS. Their composite trigger is usually big and does not meet the detection key idea of Neural Cleanse whereby the trigger is considered to be small enough. Besides, benign features won't make neurons highly activated, thus they can escape from the detection of ABS. Liu et al. TaCT proposed in \cite{b46} designs a more complex `misclassification rule' that it requires its trigger is only triggered by a specific class. Through this way, TaCT reduces its model anomalies. \cite{b40} carefully design the trigger as a mirror object after reflection to achieve the stealthiness of it and keep the classification accuracy on the clean dataset. However, not all the reflection images can be backdoor triggers. A too small reflection object is not aggressive enough while a too strong reflection object breaks the stealthiness of the backdoor. Rakin et al. \cite{b25} propose a new attack that does not need to intervene the training process of DNN. They inject backdoors by flipping some specific bits of weight parameters stored in DRAM. Their triggers are calculated to trigger corresponding bits. Besides, they need the access to the architecture of the computer. Quiring and Rieck \cite{b24} combined data poisoning and image-scaling to deploy new backdoor attack. There is no trigger in the vision until the poisoned image is downscaled. WaNet \cite{b47} uses a special warping as a trigger. It calculates a satisfied warping method to deal with all the target images. These warped images would be identified as another class. DFST \cite{b48} transfers the style of the target images and uses the specific style as an trigger. It trains a trigger generator to add the specific style for images. At the same time, it deploys detoxification to its model.
  
  Most of these attacks require a delicate trigger with too much limitation while SGBA allows the shape and pattern of the trigger to be free enough. Both of \cite{b5} and \cite{b46}, the universal of their triggers decreases. TaCT \cite{b46} can only affect one specific class. Composite attack \cite{b5} is suit for the scenarios where the two benign objects of the trigger appears at the same time. When the attacker attacks the third class, she/he needs to attach both of the two benign objects on the images which would decrease the stealthiness and the attack success rate of the trigger. Compared with them, SGBA keeps good universal of the trigger. TBT attack \cite{b25} relies on hardware while SGBA does not. As to \cite{b24}, it aims to make backdoor attacks escape from eyes of human while our target is to escape from machines. Moreover, these attacks do not take ML-based schemes into consideration.
  
  Besides, WaNet \cite{b47} and DFST \cite{b48} are hard to deploy in the reality. Although a warped image can escape from human and machine, it's hard to warp a real object in the physical world. It is also difficult for DFST to change the style of the real environments. Compared with them, the patch-based trigger of SGBA is more easy to be attached on the target object in the reality. DFST \cite{b48} takes ULP \cite{b35} into consideration, however, it doesn't expose itself to the training set of ULP \cite{b35} in which situation it tends to be escape from the detection more easily.

  \textbf{Defense:}
  Besides the mainstream model inspection schemes studied in the paper, there are some other defense schemes. Practical Detection proposed by Wang et al. \cite{b13} is also an improvement based on Neural Cleanse. Besides benign triggers for each class, it generates reversed triggers for each input. It claims a model is backdoored only when the reversed trigger of class $A$ appears similar to the reversed trigger of each single input. Although the style of generating reversed triggers is changed, the search method does not. So this scheme would still find the scapegoat instead of the real trigger when detecting SGBA models and we haven't discussed it above. We do not take undifferentiated defense schemes like Fine-Pruning \cite{b15} into consideration because they directly remove the part that could be trojaned but cannot tell whether a model is malicious or not. Besides, they always cause a loss in model accuracy. Spectral Signatures \cite{b11} and Activation Clustering \cite{b17} also do detection before model deploying. But they claim that the defense has rights to access training set and network weights of the target model. However, we suppose that the malicious third party only release the target model because there is no benefit for them to release their poisoned training set, so we do not consider them in this paper. What's more, we don't take dynamic detections that detect backdoors during the running process into consideration either. SentiNet \cite{b12} and STRIP \cite{b10} both think that the trigger in the input contributes the most to the classification result. Februus \cite{b16} protects models by clearing the trigger in the input. All of them need to check and detect every input while the model is running. We think that it will consume too much resources. There are some other special defense schemes. Some works choose to design specialized hardware to assist detection \cite{b32,b34}. Kaviani et al. \cite{b33} design a more complex constructure of DNN to protect models from backdoor attacks. Although these works are effective but they need to change the current way of DNN designation and application.

  \textbf{Other Areas:}
  This paper focuses on the computer vision area. Backdoor attacks are also widely discussed in many other areas. Zhao et al. \cite{b26} propose how to deploy backdoor attacks against a video recognition model. Besides, wireless communication \cite{b27}, federated learning system \cite{b28,b29,b31} and Internet of Things \cite{b29,b30} are concerned about backdoor attacks. We choose computer vision because this field is still the most discussed. The researches on attack and defense of this field are more comprehensive, challenging and universal.

\section{conclusion}
  We have proposed a new effective PTA through a scapegoat design and weight limitation scheme against model inspection schemes in which, backdoor detection is carried out prior to deployment. Our attack has strong stealthiness and high attack accuracy. We take experiments on $3$ popular datasets and test them with $5$ specific defense schemes. Moreover, we validate the effectiveness of our attack on a more big and complex dataset -- ImageNet. The results are satisfactory. According to the experience during our study in backdoor attacks, the possible defense scheme to protect the DNN models from SGBA is to improve the search strategies. Research on backdoors in DNN is still needed. We hope to help the attack and defense of backdoors in DNN more complete and sufficient. 

\bibliographystyle{IEEEtran}
\bibliography{main}

\newpage
\appendix

  \begin{table}[htbp]
    \centering
    \footnotesize
    \caption{Detailed architecture of MNIST classifier. \\ * means the layer is followed by a Dropout.}
    \scalebox{1.0}{
        \begin{tabular}{ccccc}
            \toprule
            Layer & Filter & Filter Size & Stride & Activation \\
            \midrule
            Conv2d* & $32$ & $5 \times 5$ & $1$ & ReLU \\
            Maxpool & $32$ & $2 \times 2$ & $2$ & - \\
            Conv2d* & $64$ & $5 \times 5$ & $1$ & ReLU \\
            Maxpool & $64$ & $2 \times 2$ & $2$ & - \\
            Linear*  & $512$ & - & - & ReLU \\
            Linear & $10$ & - & - & Softmax \\
            \bottomrule
        \end{tabular}
    }
    \label{table:mnist}
  \end{table}

  \begin{table}[hbp]
    \centering
    \footnotesize
    \caption{Detailed architecture of CIFAR10 classifier. \\ * means the layer is followed by a Dropout.}
    \scalebox{1.0}{
        \begin{tabular}{ccccc}
            \toprule
            Layer & Filter & Filter Size & Stride & Activation \\
            \midrule
            Conv2d & $32$ & $3 \times 3$ & $1$ & ReLU \\
            Conv2d & $32$ & $3 \times 3$ & $1$ & ReLU \\
            Maxpool & $32$ & $2 \times 2$ & $2$ & - \\
            Conv2d & $64$ & $3 \times 3$ & $1$ & ReLU \\
            Conv2d & $64$ & $3 \times 3$ & $1$ & ReLU \\
            Maxpool & $64$ & $2 \times 2$ & $2$ & - \\
            Linear  & $256$ & - & - & ReLU \\
            Linear*  & $256$ & - & - & ReLU \\
            Linear & $10$ & - & - & Softmax \\
            \bottomrule
        \end{tabular}
    }
  \label{table:cifar10}
  \end{table}

  \begin{table}[hbp]
    \centering
    \footnotesize
    \caption{Detailed architecture of GTSRB classifier. \\ * means the layer is followed by a Dropout.}
    \scalebox{1.0}{
        \begin{tabular}{ccccc}
            \toprule
            Layer & Filter & Filter Size & Stride & Activation \\
            \midrule
            Conv2d & $32$ & $3 \times 3$ & $1$ & ReLU \\
            Conv2d & $32$ & $3 \times 3$ & $1$ & ReLU \\
            Maxpool* & $32$ & $2 \times 2$ & $2$ & - \\
            Conv2d & $64$ & $3 \times 3$ & $1$ & ReLU \\
            Conv2d & $64$ & $3 \times 3$ & $1$ & ReLU \\
            Maxpool* & $64$ & $2 \times 2$ & $2$ & - \\
            Conv2d & $128$ & $3 \times 3$ & $1$ & ReLU \\
            Conv2d & $128$ & $3 \times 3$ & $1$ & ReLU \\
            Maxpool* & $128$ & $2 \times 2$ & $2$ & - \\
            Linear*  & $512$ & - & - & ReLU \\
            Linear & $43$ & - & - & Softmax \\
            \bottomrule
        \end{tabular}
    }
    \label{table:gtsrb}
  \end{table}

  \begin{table*}[hbp]
    \centering
    \footnotesize
    \caption{Detailed architecture of ImageNet classifier. \\ ReLU is the default activation method.}
    \scalebox{1.0}{
        \begin{tabular}{c|c|c}
            \toprule
            \multicolumn{2}{c|}{Layer} & Structure \\
            \midrule
            \multicolumn{2}{c|}{conv1} & Conv2d(3, 64, kernel\_size=(7, 7), stride=(2, 2), padding=(3, 3), bias=False) \\ \hline
            \multicolumn{2}{c|}{bn1} & BatchNorm2d(64, eps=1e-05, momentum=0.1, affine=True, track\_running\_stats=True \\ \hline
            \multicolumn{2}{c|}{maxpool} & MaxPool2d(kernel\_size=3, stride=2, padding=1, dilation=1, ceil\_mode=False) \\ \hline
            \multirow{15}*{layer1} & \multirow{8}*{Bottleneck1} & Conv2d(64, 64, kernel\_size=(1, 1), stride=(1, 1), bias=False)\\
            \cline{3-3}
            & & BatchNorm2d(64, eps=1e-05, momentum=0.1, affine=True, track\_running\_stats=True) \\
            \cline{3-3}
            & & Conv2d(64, 64, kernel\_size=(3, 3), stride=(1, 1), padding=(1, 1), bias=False) \\
            \cline{3-3}
            & & BatchNorm2d(64, eps=1e-05, momentum=0.1, affine=True, track\_running\_stats=True) \\
            \cline{3-3}
            & & Conv2d(64, 256, kernel\_size=(1, 1), stride=(1, 1), bias=False) \\
            \cline{3-3}
            & & BatchNorm2d(256, eps=1e-05, momentum=0.1, affine=True, track\_running\_stats=True) \\
            \cline{3-3}
            & & \makecell{downsample: Sequential((0): Conv2d(64, 256, kernel\_size=(1, 1), stride=(1, 1), bias=False) \\
            (1): BatchNorm2d(256, eps=1e-05, momentum=0.1, affine=True, track\_running\_stats=True))} \\
            \cline{2-3}
            & \multirow{6}*{Bottleneck2} & Conv2d(256, 64, kernel\_size=(1, 1), stride=(1, 1), bias=False) \\
            \cline{3-3}
            & & BatchNorm2d(64, eps=1e-05, momentum=0.1, affine=True, track\_running\_stats=True) \\
            \cline{3-3}
            & & Conv2d(64, 64, kernel\_size=(3, 3), stride=(1, 1), padding=(1, 1), bias=False) \\
            \cline{3-3}
            & & BatchNorm2d(64, eps=1e-05, momentum=0.1, affine=True, track\_running\_stats=True) \\
            \cline{3-3}
            & & Conv2d(64, 256, kernel\_size=(1, 1), stride=(1, 1), bias=False)\\
            \cline{3-3}
            & & BatchNorm2d(256, eps=1e-05, momentum=0.1, affine=True, track\_running\_stats=True)\\
            \cline{2-3}
            & Bottleneck3 & the same as Bottleneck2\\
            \hline
            \multirow{15}*{layer2} & \multirow{8}*{Bottleneck1} & Conv2d(256, 128, kernel\_size=(1, 1), stride=(1, 1), bias=False) \\
            \cline{3-3}
            & & BatchNorm2d(128, eps=1e-05, momentum=0.1, affine=True, track\_running\_stats=True)\\
            \cline{3-3}
            & & Conv2d(128, 128, kernel\_size=(3, 3), stride=(2, 2), padding=(1, 1), bias=False)\\
            \cline{3-3}
            & & BatchNorm2d(128, eps=1e-05, momentum=0.1, affine=True, track\_running\_stats=True)\\
            \cline{3-3}
            & & Conv2d(128, 512, kernel\_size=(1, 1), stride=(1, 1), bias=False)\\
            \cline{3-3}
            & & BatchNorm2d(512, eps=1e-05, momentum=0.1, affine=True, track\_running\_stats=True)\\
            \cline{3-3}
            & & \makecell{downsample: Sequential((0): Conv2d(256, 512, kernel\_size=(1, 1), stride=(2, 2), bias=False)\\
            (1): BatchNorm2d(512, eps=1e-05, momentum=0.1, affine=True, track\_running\_stats=True))}\\
            \cline{2-3}
            & \multirow{6}*{Bottleneck2} & Conv2d(512, 128, kernel\_size=(1, 1), stride=(1, 1), bias=False)\\
            \cline{3-3}
            & & BatchNorm2d(128, eps=1e-05, momentum=0.1, affine=True, track\_running\_stats=True)\\
            \cline{3-3}
            & & Conv2d(128, 128, kernel\_size=(3, 3), stride=(1, 1), padding=(1, 1),
            bias=False)\\
            \cline{3-3}
            & & BatchNorm2d(128, eps=1e-05, momentum=0.1, affine=True, track\_running\_stats=True)\\
            \cline{3-3}
            & & Conv2d(128, 512, kernel\_size=(1, 1), stride=(1, 1), bias=False)\\
            \cline{3-3}
            & & BatchNorm2d(512, eps=1e-05, momentum=0.1, affine=True, track\_running\_stats=True)\\
            \cline{2-3}
            & Bottleneck3 & the same as Bottleneck2 \\
            \cline{2-3}
            & Bottleneck4 & the same as Bottleneck2 \\
            \hline
            \multirow{18}*{layer3} & \multirow{8}*{Bottleneck1} & Conv2d(512, 256, kernel\_size=(1, 1), stride=(1, 1), bias=False)\\
            \cline{3-3}
            & & BatchNorm2d(256, eps=1e-05, momentum=0.1, affine=True, track\_running\_stats=True)\\
            \cline{3-3}
            & & Conv2d(256, 256, kernel\_size=(3, 3), stride=(2, 2), padding=(1, 1),
            bias=False)\\
            \cline{3-3}
            & & BatchNorm2d(256, eps=1e-05, momentum=0.1, affine=True, track\_running\_stats=True)\\
            \cline{3-3}
            & & Conv2d(256, 1024, kernel\_size=(1, 1), stride=(1, 1), bias=False)\\
            \cline{3-3}
            & & BatchNorm2d(1024, eps=1e-05, momentum=0.1, affine=True, track\_running\_stats=True)\\
            \cline{3-3}
            & & \makecell{downsample: Sequential((0): Conv2d(512, 1024, kernel\_size=(1, 1), stride=(2, 2), bias=False)\\
            (1): BatchNorm2d(1024, eps=1e-05, momentum=0.1, affine=True, track\_running\_stats=True))}\\
            \cline{2-3}
            & \multirow{6}*{Bottleneck2} & Conv2d(1024, 256, kernel\_size=(1, 1), stride=(1, 1), bias=False)\\
            \cline{3-3}
            & & BatchNorm2d(256, eps=1e-05, momentum=0.1, affine=True, track\_running\_stats=True)\\
            \cline{3-3}
            & & Conv2d(256, 256, kernel\_size=(3, 3), stride=(1, 1), padding=(1, 1),
            bias=False)\\
            \cline{3-3}
            & & BatchNorm2d(256, eps=1e-05, momentum=0.1, affine=True, track\_running\_stats=True)\\
            \cline{3-3}
            & & Conv2d(256, 1024, kernel\_size=(1, 1), stride=(1, 1), bias=False)\\
            \cline{3-3}
            & & BatchNorm2d(1024, eps=1e-05, momentum=0.1, affine=True, track\_running\_stats=True)\\
            \cline{2-3}
            & Bottleneck3 & the same as Bottleneck2\\
            \cline{2-3}
            & Bottleneck4 & the same as Bottleneck2\\
            \cline{2-3}
            & Bottleneck5 & the same as Bottleneck2\\
            \cline{2-3}
            & Bottleneck6 & the same as Bottleneck2\\
            \hline
            \multirow{15}*{layer4} & \multirow{8}*{Bottleneck1} & Conv2d(1024, 512, kernel\_size=(1, 1), stride=(1, 1), bias=False)\\
            \cline{3-3}
            & & BatchNorm2d(512, eps=1e-05, momentum=0.1, affine=True, track\_running\_stats=True)\\
            \cline{3-3}
            & & Conv2d(512, 512, kernel\_size=(3, 3), stride=(2, 2), padding=(1, 1), bias=False)\\
            \cline{3-3}
            & & BatchNorm2d(512, eps=1e-05, momentum=0.1, affine=True, track\_running\_stats=True)\\
            \cline{3-3}
            & & Conv2d(512, 2048, kernel\_size=(1, 1), stride=(1, 1), bias=False)\\
            \cline{3-3}
            & & BatchNorm2d(2048, eps=1e-05, momentum=0.1, affine=True, track\_running\_stats=True)\\
            \cline{3-3}
            & & \makecell{downsample: Sequential((0): Conv2d(1024, 2048, kernel\_size=(1, 1), stride=(2, 2), bias=False)\\
            (1): BatchNorm2d(2048, eps=1e-05, momentum=0.1, affine=True, track\_running\_stats=True))}\\
            \cline{2-3}
            & \multirow{6}*{Bottleneck2} & Conv2d(2048, 512, kernel\_size=(1, 1), stride=(1, 1), bias=False)\\
            \cline{3-3}
            & & BatchNorm2d(512, eps=1e-05, momentum=0.1, affine=True, track\_running\_stats=True)\\
            \cline{3-3}
            & & Conv2d(512, 512, kernel\_size=(3, 3), stride=(1, 1), padding=(1, 1),
            bias=False)\\
            \cline{3-3}
            & & BatchNorm2d(512, eps=1e-05, momentum=0.1, affine=True, track\_running\_stats=True)\\
            \cline{3-3}
            & & Conv2d(512, 2048, kernel\_size=(1, 1), stride=(1, 1), bias=False)\\
            \cline{3-3}
            & & BatchNorm2d(2048, eps=1e-05, momentum=0.1, affine=True, track\_running\_stats=True)\\
            \cline{2-3}
            & Bottleneck3 & the same as Bottleneck2 \\
            \hline
            \multicolumn{2}{c|}{avgpool} & AdaptiveAvgPool2d(output\_size=(1, 1))\\
            \hline
            \multicolumn{2}{c|}{\multirow{3}*{fc}} & Linear(in\_features=2048, out\_features=256, bias=True)\\
            \cline{3-3}
            \multicolumn{2}{c|}{} & Dropout(p=0.4, inplace=False)\\
            \cline{3-3}
            \multicolumn{2}{c|}{} & Linear(in\_features=256, out\_features=50, bias=True)\\
            \bottomrule
        \end{tabular}
    }
  \label{table:imagenet}
  \end{table*}

\end{document}